\documentclass[aps,pre,onecolumn,showpacs,amsmath,amssymb,longbibliography,notitlepage,11pt,tightenlines]{revtex4-1}
\usepackage{graphicx} 
\usepackage{dcolumn} 
\usepackage{bm} 
\usepackage{xcolor}
\RequirePackage{lettrine} 
\newcommand{\dropcapfont}{\fontfamily{lmss}\bfseries\fontsize{26pt}{28pt}\selectfont}
\newcommand{\dropcap}[1]{\lettrine[lines=2,lraise=0.05,findent=0.1em, nindent=0em]{{\dropcapfont{#1}}}{}}
\usepackage[normalem]{ulem}  
\newcommand{\beginsupplement}{%
        \setcounter{table}{0}
        \renewcommand{\thetable}{S\arabic{table}}%
        \setcounter{figure}{0}
        \renewcommand{\thefigure}{S\arabic{figure}}%
     }
\begin{document}

\title{Small-scale demixing in confluent biological tissues}

\author{Preeti Sahu\textit{$^{1,*}$}, Daniel M. Sussman\textit{$^{1,2}$}, Matthias R{\"u}bsam\textit{$^3$}, Aaron F. Mertz\textit{$^{4}$}, Valerie Horsley\textit{$^{5}$}, Eric R. Dufresne\textit{$^{4,6,7}$}, Carien M. Niessen\textit{$^3$}, M. Cristina Marchetti\textit{$^8$},
M. Lisa Manning\textit{$^1$}, J. M. Schwarz\textit{$^{1,9*}$}}
\email{prsahu@syr.edu; $^{**}$ jschwarz@physics.syr.edu}
\affiliation{$^1$Department of Physics and BioInspired Syracuse: Institute for Material and Living Systems, Syracuse University, Syracuse, NY 13244, USA\\
$^2$ Department of Physics, Emory University, Atlanta, GA, 30322, USA\\
$^3$ Department of Dermatology, CECAD Cologne, Center for Molecular Medicine, University of Cologne, Cologne, Germany\\
$^4$ Department of Physics, Yale University, New Haven, CT 06520, USA\\
$^5$ Department of Molecular, Cellular and Developmental Biology, Yale University, New Haven, CT 06520, USA\\
$^6$ Departments of Mechanical Engineering and Materials Science, Chemical and Environmental Engineering, and Cell Biology, Yale University, New Haven, CT, 06520, USA\\
$^7$ Department of Materials, ETH Z{\"u}rich, 8093 Z{\"u}rich, Switerland\\
$^8$Department of Physics, University of California at Santa Barbara, Santa Barbara, CA 93106, USA\\
$^9$ Indian Creek Farm, Ithaca, NY 14850, USA}

\date{\today}

\begin{abstract}
{Surface tension governed by differential adhesion can drive fluid particle mixtures to sort into separate regions, i.e., demix. Does the same phenomenon occur in confluent biological tissues? We begin to answer this question for epithelial monolayers with a combination of theory via a vertex model and experiments on keratinocyte monolayers. Vertex models are distinct from particle models in that the interactions between the cells are shape-based, as opposed to distance-dependent. We investigate whether a disparity in cell shape or size alone is sufficient to drive demixing in bidisperse vertex model fluid mixtures. Surprisingly, we observe that both types of bidisperse systems robustly mix on large lengthscales. On the other hand, shape disparity generates slight demixing over a few cell diameters, a phenomenon we term micro-demixing. This result can be understood by examining the differential energy barriers for neighbor exchanges (T1 transitions). Experiments with mixtures of wild-type and E-cadherin-deficient keratinocytes on a substrate are consistent with the predicted phenomenon of micro-demixing, which biology may exploit to create subtle patterning. The robustness of mixing at large scales, however, suggests that despite some differences in cell shape and size, progenitor cells can readily mix throughout a developing tissue until acquiring means of recognizing cells of different types.}
\end{abstract}

\maketitle

\bigskip
\dropcap{L}iquid-liquid phase separation, i.e., demixing, drives patterning. In materials science, demixing between two liquids is typically driven by the energetics of interfacial tension overcoming entropy-driven mixing~\cite{Safran2002}. By cooling a material, one can tune between a mixed state at high temperature and a demixed state at low temperature. Depending on the material and quench rate, this transition can occur continuously via spinodal decomposition or discontinuously via nucleation~\cite{FrenkelLouis,CahnHillard}. In order to distinguish between mechanisms it is often useful to analyze the lengthscales of emergent patterns: nucleation and spinodal decomposition give rise to characteristic lengthscales that then coarsen, while in the absence of interfacial tension fluids will mix down to the scale of single molecules.
These and related demixing phenomena have been studied numerically using multicomponent Lennard-Jones mixtures in which particles have a fixed shape and an interaction potential that depends on the distance between. The potential also energetically distinguishes between particles of different types to model interfacial tension~\cite{RowlinsonSwinton}. 

In biology, demixing at the subcellular scale can lead to compartmentalization within cells~\cite{Feric2016}, while in a developing organism, demixing can lead to compartmentalization among cells of different type, otherwise known as cell sorting. In fact, interfacial tension-driven demixing has long been invoked to explain cellular patterning. The first among such ideas is the Differential Adhesion Hypothesis (DAH), proposed by Steinberg in 1963~\cite{STEINBERG1963}, to explain patterns in the spatial sorting of progenitor cells, such as ectoderm and mesoderm, during embryonic development. The DAH postulates that tissues behave like immiscible liquids composed of motile cells that rearrange in order to minimize their interfacial tension caused by differences in cell-cell adhesion. Building on the DAH, Harris~\cite{Harris1976a}, and later Brodland~\cite{Brodland2002}, have highlighted the importance of other contributors to interfacial tension, including regulation of the acto-myosin cortex. There is an emerging consensus~\cite{Yamada2007, Maitre2012, Manning2010a, Amack2012, Mertz2012} that adhesive molecules help to regulate cortical acto-myosin, which can strongly impact cell sorting. However, it remains controversial whether differential adhesion or differential cortical tension alone is sufficient to generate the level of cell sorting and compartmentalization observed in embryos and cell culture systems~\cite{Heasman1993,Song2016,Ninomiya2012,Pawlizak2015,Dong2018,Landsberg2009,Cochet-Escartin2017}. Several experiments have suggested that additional processes such as specialized cell-cell signaling~\cite{Ninomiya2012} or cellular jamming~\cite{Pawlizak2015} enhance or disrupt sorting in living tissues.

One major difference between immiscible liquids composed of cells and immiscible liquids composed of soft spheres is that in the latter case, the particles have a distance-dependent interaction, while in epithelial layers and even some 3D tissues, the cells are confluent -- they can change their shape to completely fill space---and so their interaction is shape-based. To reflect this property, confluent tissues have been studied theoretically and computationally using vertex or Voronoi models~\cite{Nagai2001,Farhadifar2007b,Staple2010a,Barton2017}, where cells are constructed from tessellations of space with no gaps between cells. As active fluctuations drive cellular rearrangements, cells must deform so that no gaps open up between them. This suggests cells are subject to strong geometrical and topological constraints. For example, in flat 2D tilings with three-fold coordinated vertices, the average number of neighbors must be precisely six. This constraint leads one to predict that a rigidity transition should occur when neighbor exchange between six-sided cells cost zero energy, i.e. when cells can form regular pentagons at zero cost~\cite{Bi2015g,Bi2016a}. This prediction has since been realized in experiments~\cite{Park2015a} and is distinct from rigidity transitions in particulate systems~\cite{SM2018,MBM2018}. 

Does such an interaction potential with non-trivial geometrical and topological constraints affect the fundamental definition of surface tension? Work on bidisperse foams modeled as ordered vertex models demonstrate that, in equilibrium, demixed cells of two different areas have a lower energy than a mixed system and so demixed states are energetically preferred~\cite{Graner2002}. However, disperse-in-area foams under large shear strain will mix~\cite{Cox2006}. If we think of the shear strain as a temperature-like variable, then these findings are similar to particulate systems.  

On the other hand, recent work by some of us demonstrates that so-called heterotypic contacts in vertex models can drastically affect the notion of interfacial tension~\cite{Sussman2018b}. Heterotypic contacts, where cells recognize neighbors of a different cell type, can be modeled in two-dimensional vertex models with a higher or lower line tension along interfaces between cells of different types, or heterotypic line tension. Such a rule results in very sharp, yet deformable, interfaces~\cite{Sussman2018b} where surface tension measured by macroscopic deformation of an overall droplet shape gives a value in line with equilibrium expectations, yet, surface tension measured from interfacial fluctuations is at least an order of magnitude larger. This discrepancy is due to discontinuous pinning forces generated during topological rearrangements between cells of different types. That is, it is a consequence of the shape-based nature of the interactions. 

Here, we explore the possibility of interfacial-tension-driven demixing in the absence of explicit heterotypic tension in both modeling and in experiments. From the modeling side, we consider a two-dimensional vertex model with two different cell types. Particulate mixtures can demix when a miscibility parameter, the ratio of the strength of the distance-dependent interaction between dissimilar particles as compared to similar particles, becomes less than one. Since in vertex models the interaction is shape-based, it is natural to ask if binary vertex model fluids consisting of mixtures of cells with different preferred cell shapes and/or sizes, accounting for differential adhesion, cortical tension or volume, demix even in the absence of specialized heterotypic interactions. In other words, is there an emergent effective interfacial tension between two cell types that is sufficient enough to sort cells? Should the answer be yes, then one can imagine that the sorting of progenitor cells occurs very early on in the development process before robust heterotypic interfacial tensions are established. Should the answer be no, then cells must establish heterotypic interactions before sorting can occur, suggesting a more important mechanical role for cell recognition receptors than previously thought. The topological nature of the discontinuous pinning forces stabilizing interfaces in vertex model fluid mixtures tells us that once such recognition is in place, a finite active force is required to overcome the discontinuity~\cite{Sussman2018b}. Interestingly, a recent study with both {\it in vitro} experiments and cellular Potts model simulations suggests that a large heterotypic line tension is required for cell sorting~\cite{Canty2017b}, although the mechanism was left unresolved.

In searching for whether or not large-scale interfacial tensions and, therefore, cell demixing are emergent/collective properties of such binary vertex model fluid mixtures, we do not observe large-scale demixing. However, we do observe small-scale demixing in mixtures with differential adhesion, which is not thermodynamic in origin and which we term micro-demixing. We find that this behavior arises from dynamical trapping due to energy barriers to neighbor exchanges (T1 transitions) that depend on configuration of the type of cell.  We then ask if the predicted phenomenon of micro-demixing can be realized in cellular systems.  

To begin to answer this question, we experimentally study monolayers of mixtures of wild-type primary keratinocytes, denoted as Ctr cells, and primary keratinocytes in which the E-cadherin has been knocked down, denoted as E-cad$^{-/-}$ cells. Cadherins, such as E-cadherin and P-cadherin, are crucial components of adherens junctions (AJs) that couple intercellular adhesion to the cytoskeletin via $\alpha$- and $\beta$-catenin~\cite{cadherin}, the former of which can interact directly with actin and other actin binding proteins. It been even more recently established that E-cadherin plays a central role in the mechanical circuitry coordinating adhesion, contractile forces and biochemical signaling to drive polarized organization of tension observed in stratified epidermal layers~\cite{Rubsam}.  Given the central role of E-cadherin, E-cad$^{-/-}$ keratinocytes affect the mechanical circuitry via, for example, decreased adhesion site lengths~\cite{Rubsam}. Since the wild-type keratinocytes contain both P- and E-cadherin, two-dimensional mixtures of the wild-type keratinocytes with E-cad$^{-/-}$ keratinocytes are ideal for testing whether or not differential adhesion leads to large-scale demixing or not, even in the absence of heterotypic tensions. In addition to the lack of large-scale demixing in the experiments, we also find evidence for small-scale demixing as predicted in our two-dimensional vertex model, further bolstering the use of this class of models as a predictor of tissue rheology.

\bigskip
\noindent\textbf{\large Computational Model}\\

Cells are biomechanical (and biochemical) constructs that are not in equilibrium, i.e. they are driven by active forces. Given our question of mixing, we study a confluent monolayer of cells of different types. The biomechanics of the $j$th cell of type $\beta$ is given by the energy functional:
      \begin{equation}
      \label{eq:energy}
      E_{j,\beta}= K_a (A_{j,\beta} - A_{0,\beta})^2+K_p (P_{j,\beta} -P_{0,\beta})^2, 
      \end{equation}
      where $A_{j,\beta}$ denotes the $j$th cell area of type $\beta$ and the $j$th cell perimeter of type $\beta$ is denoted by $P_{j,\beta}$. Given the quadratic penalty from deviating for a cell's preferred area and perimeter, $K_a$ and $K_p$ are area and perimeter stiffnesses, respectively, and both are independent of cell type. Physically, the area term represents the bulk elasticity of the cell, while the perimeter represents the contractility of the acto-myosin cortex with $P_{0,\beta}$ denoting a competition between cortical tension and cell-cell adhesion. The total energy of the tissue is then defined as $E=\sum_{j,\beta} E_{j,\beta}$. An important parameter in these models is the dimensionless shape index $s_{0,\beta} = P_{0,\beta}/ \sqrt{A_{0,\beta}} $. A regular hexagon has a dimensionless shape index of $s_0\approx 3.72$, for example. 
      
\begin{figure}[h!]
\centering
\includegraphics[width=0.99\textwidth]{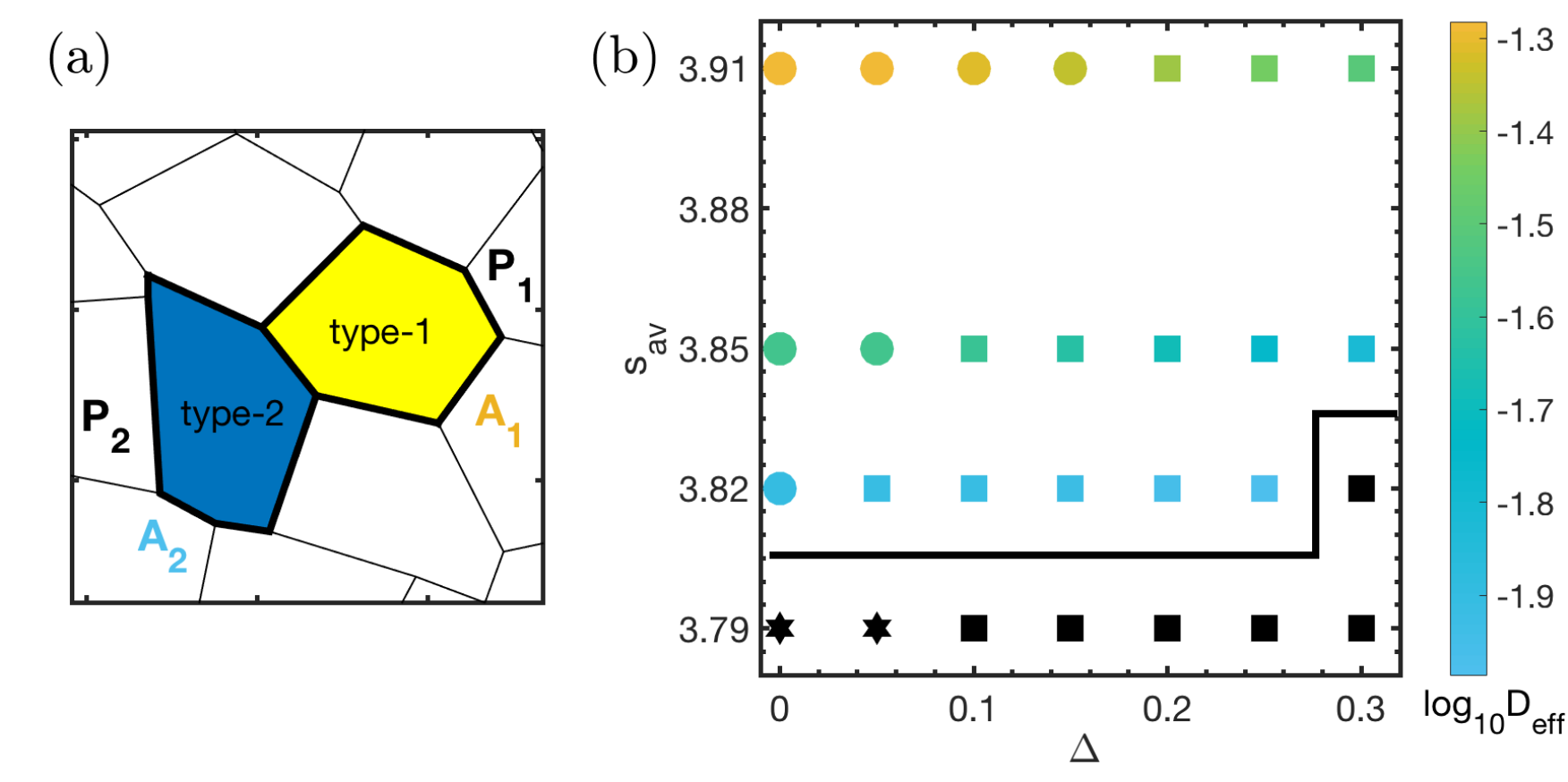}
\caption{{\it Vertex model binary mixtures.} (a) Schematic of vertex-based modeling of a tissue: A typical tessellation with two different types of cells highlighted. The energy depends only on a cell's perimeter ($P_j$) and area ($A_j$). (b) A heat map of $\log_{10}D_{eff}$ as a function of $\Delta$ on the x-axis and $s_{av}$ on the y-axis. The phase points with: fluid-fluid ($s_{0,1},s_{0,2}>3.81$), solid-fluid ($s_{0,1}<3.81$) and solid-solid ($s_{0,1},s_{0,2}<3.81$) components are denoted by circular, square and star-shaped markers, respectively. Black-filled markers, demarcated by a solid black line, denote mixtures with a $D_{eff}$ less than that of the chosen cutoff of $0.01$. Region above this line denotes fluid-like behaviour on average. }
\label{fig:MSD}
\end{figure}

To study binary mixtures, we fix $\beta=1,2$ and allow the cell types to have different parameters, $A_{0,\beta}$ and $s_{0,\beta}$ (see Fig.~\ref{fig:MSD}a). What is the biological implication of two different shape indices, for example? Consider two cell-types with the exact same area. The cell type that prefers to a have higher shape index, can do so by increasing the density of adhesion molecules, for example. Therefore, mixing these two differently adhering cell-types corresponds to studying mixing in cell-types with two different shapes. In reality, these adhesion receptors also affect the cell shape by signaling to either up-regulate or down-regulate contractility. In the vertex model energy functional that we use, we have packaged both adhesion and contractility into the preferred shape index. Though more detailed models are possible, we find that our minimal model does indeed tell us something about how cells behave as indicated by the experiments presented below.  We will focus on cases of 50:50 mixtures where there is an equal number of each cell type, with either preferred shape disparity or preferred area disparity. Unless otherwise specified, the two components are uniformly distributed in the initial state. We set $K_p$ to unity for all systems. 

We study the above energy functional from an energy minimization perspective as well as from a dynamical perspective in which the cells migrate within the monolayer.  As for the latter, there is still much debate about how to model the motility of cells.  We have chosen to model the motility of a cell by imposing a random active force on each vertex, i.e. each vertex undergoes over-damped Brownian motion at a fixed effective temperature $T$ with a conservative force contribution from the above energy functional and a second force contribution from a Brownian force. While there are other possible dynamical rules, we have found that, for example, the properties of an interface between two cells types with heterotypic line tension betwen them is rather robust to the specifics of the dynamical rule~\cite{Sussman2018b}. The equation of motion for each vertex is then iterated until the cells can adequately explore the entire system such that the system approaches a steady state, at least for most parameters we study. As the cells move, they may rearrange and come into contact with new cells. Such rearrangements are known as T1 transitions. To implement a T1 transition, an edge shared between two cells undergoes a $\pi/2$ rotation once the length of this edge is below some threshold length.  The rotated edge then lengthens and allows for two different cells to now share an edge. As for the energy minimization approach, in addition to comparing the minimum energy configurations of both demixed and mixed states, we will also compute the energy barriers associated with T1 transitions by constraining the length of a particular edge in the system such that a T1 transition occurs while allowing the remaining degrees of freedom in the system to relax.  See the Methods section for more details.

We are also interested in comparing the behavior of these bidisperse systems to ones with an explicit heterotypic line tension (HLT), where cell types $1$ and $2$ recognize their joint interface as a heterotypic interface and, therefore, alter the line tension at that interface.  Such interactions are common in cellular Potts models~\cite{Graner1992, Canty2017b} and have also been studied in vertex and Voronoi models~\cite{Barton2017,Sussman2018b}.  In this case, we add an extra term to the cell energy to arrive at:     
      \begin{equation}
        \label{eq:gamma}
        E_{HLT}  =  \sum_{j,\beta} K_a (A_{j,\beta} - A_{0,\beta})^2+K_p (P_{j,\beta} -P_{0,\beta})^2  + \gamma \sum_{\langle i,j\rangle} (1-\delta_{\alpha\beta}) l_{ij}.  
      \end{equation}
The latter sum is over all edges, $l_{ij}$, between cells $i$ and $j$ with $\delta{\alpha \beta}$ representing a Kronecker delta such that there is additional line tension only between cells of two different types $\alpha$ and $\beta$.  For simplicity, we assume that the additional tension, $\gamma$, is the same for all heterotypic edges.

\bigskip
\noindent\textbf{\large Computational Results}\\
\noindent\textbf{Stability and fluidity of shape bidisperse mixtures.}\label{rheo}
To single out the effect of shape dispersity, we first vary the preferred shapes under the constraint that the preferred/target area is the same across cell types, $A_{0,1}=A_{0,2}=1$. 

Previous work on the vertex model has identified a regime in parameter space dominated by a coarsening instability~\cite{Farhadifar2007b,Staple2010a}, where some cells shrink in size and others grow. We expect that heterogeneous $s_0$ values might amplify this instability, as heterogeneity amplifies differences between the cells. To prevent area dispersity from affecting the results in these mixtures, we choose $K_a=100$, which is sufficient to reduce fluctuations in area $A$ from target area $A_0$ to a standard deviation of less than 1 \%, preventing the onset of the coarsening instability. Moreover, Fig.~\ref{fig:highKa} shows that increased area stiffness does not significantly impact the fluidity of homogeneous tissues, as measured by the effective diffusivity (Eqs.~\ref{eq:msdDef}-\ref{eq:Def}), denoted as $D_{eff}$, which is the ratio of the diffusion constant in the presence of interactions to that in the absence of interactions. The onset of a finite effective diffusivity as a function of the shape index remains near $s_0\approx 3.81$ with increasing $K_a$.  

Next, we investigate how shape disparity affects the fluidity of the tissue. We find that the most effective way to represent the phase space of the two-component system with two shape indices $s_{0,1},s_{0,2}$ is by the average value of the shape index,  $s_{av}= (s_{0,1}+ s_{0,2})/2$, and the shape disparity, which is the difference between the two values, $\Delta=s_{0,2}-s_{0,1}$ with $s_{0,2}>s_{0,1}$. Figure~\ref{fig:MSD}b is a heat map of the effective diffusivity of binary mixtures as a function of $s_{av}$ and $\Delta$. We see that there is a boundary between fluid-like and solid-like, demarcated by the thick solid line, as determined by the $D_{eff}$ threshold. Interestingly, for $\Delta=0.3$, this boundary does not match up with the fluid-solid boundary for the monodisperse case for the Brownian limit of a self-propelled Voronoi model at a similar temperature, which is near $s_0=3.81$~\cite{Bi2016a}. Moreover, the solid-fluid mixtures depicted by squares, have a fluid-like diffusivity above the boundary line. This indicates that the fluid-like species in the mixture are sufficient to fluidize the entire tissue, which is additionally confirmed by analyzing the diffusivities of each component (Fig.~\ref{fig:CompDeff}a).

\begin{figure}[!t]
\centering
\includegraphics[width=0.99\textwidth]{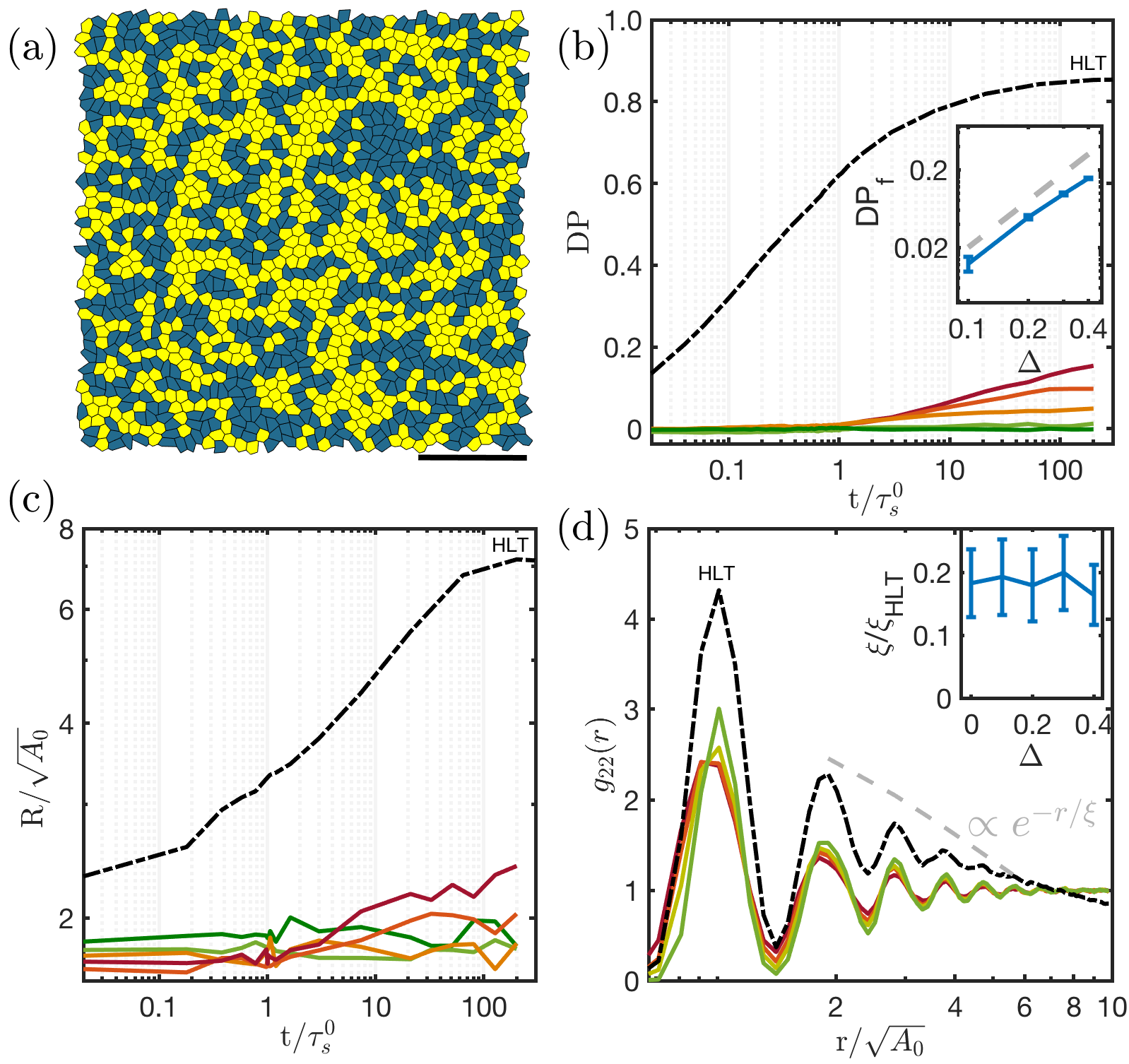}
\caption{{\it Shape bidisperse fluid mixtures.} (a) Snapshot of a $P_{av}=3.85$, $\Delta=0.4$, $N=1600$ mixture. Scale bar denotes 10 length units. Yellow is used for solid-like cells ($s_0=3.65$) and blue for liquid-like ones ($s_0=4.05$). (b)-(d) Various quantifications of demixing in shape bidisperse mixtures (curves colored from green to red in increasing order of shape disparity i.e $\Delta=0.01,0.1,0.2,0.3,0.4$) are compared to a mixture with an extra heterotypic line tension of value 0.1, $s_0=3.97$ (black dashed curve). (b) Demixing Parameter versus $\log(time)$. The final value ($DP_f$) as a function of $\Delta$ is shown in the inset. (c) Average cluster radius (R) versus time. (d) Pair correlation function of high-$s_0$ cells $(g_{22})$ versus radial distance for  $t=200\,\tau_s^{0}$. The dashed grey line shows an exponential decay. The inset shows the decay lengthscale ($\xi$) in terms of the maximum possible lengthscale ($\xi_{HLT}$) with increasing disparity $\Delta$. Simulation details provided in Table S1. 
}
{\label{fig:shapeG}}
\end{figure}

\noindent\textbf{{Binary mixtures with two target shapes.}}
After understanding how $\Delta$ affects diffusivity in a mixture, let us now understand its role in bulk demixing for a fixed $s_{av}=3.85$. A snapshot of a typical long-time configuration for such a mixture is shown in Fig.~\ref{fig:shapeG}a. By eye, it appears that demixing does occur at very small scales, due to some clustering of the cells with larger $s_0$. The system maintains this small-scale structure at long times. No large-scale demixing is observed. Hence, we shall refer to this process as {\it micro-demixing}. To quantify micro-demixing and highlight its long-time steady state, we study three observables.

The first is the demixing parameter $DP$, which measures the average environment of each species, quantifying whether it is more likely to be surrounded by similar (homotypic) or dissimilar (heterotypic) cells.  Defining $N_s$ as the number of similar neighboring cells and $N_t$ as the total number of neighboring cells, 
\begin{equation}
 \label{eq:dp}
DP=\big \langle DP_j \big \rangle = \bigg \langle 2\bigg(\frac{N_s}{N_t} - \frac{1}{2} \bigg) \bigg \rangle,
\end{equation}
where the brackets denote averaging over all cells in the tessellation. In a completely mixed state, $DP=0$, whereas in a completely sorted mixture, $DP=1$, in the limit of infinite system size.

The demixing parameter as a function of time is shown in Fig.~\ref{fig:shapeG}b. The value of DP is initially zero since the two cell types are initially seeded at random, and saturates to a small non-zero value at long times. The final steady state value, $DP_f$, increases with increasing shape disparity $\Delta$, as shown in the inset to Fig.~\ref{fig:shapeG}b, and the length of time required to reach the steady state also increases with increasing $\Delta$ (Fig. \ref{fig:CompDeff}(b)).  
For comparison, the dashed black line in Fig.~\ref{fig:shapeG}b illustrates the demixing parameter as a function of time for a model with heterotypic line tension. In the HLT case, DP rises very quickly to a value close to unity as one species rapidly forms a circular droplet, in a manner similar to that expected for conventional liquid-liquid binary mixtures.  

We then measure the average cluster radius $R$ by quantifying the average radius of gyration of the dispersed component. In the case of shape bidisperse mixtures, the more fluid-like (larger $s_0$) component tends to be dispersed. The average cluster radius (Fig.~\ref{fig:shapeG}c) shows a small growth in time, which appears to saturate at long times, although the data is noisier given the cluster statistics sampling rate. The steady state radius tends to increase with increasing $\Delta$. In all cases studied, clusters have an average radius of less than $2.5 \pm 0.2$. For comparison, the dashed line shows a system with HLT, which we expect to saturate as a nearly circular droplet of one species embedded inside the other species. For the system size we study, this would correspond to a cluster of radius $~8$, which is close to the observed steady state value of $7.2\pm 0.2$.

To further quantify the structure of this micro-demixed state, we study the pair correlation function, $g(r)$, which describes the normalized probability of finding a cell center a given distance from another cell center.  In homogeneous fluids and amorphous solids, this function exhibits short range order with peaks occurring at distances that are integer multiples of the typical spacing between two cells.  The envelope of these peaks falls off with distance and eventually approaches unity, highlighting that these materials are disordered over larger lengthscales. In bidisperse mixtures, we compute the correlation between each species $\beta$ separately, defined by the relative position vectors ($r^{(\beta)}$) between two cells of type $\beta$:
\begin{equation}
 \label{eq:gr}
g_{\beta\beta}(r) =\frac{1}{2\pi r N_{\beta} \rho_0} \sum\limits_i^{N_{\beta}}\sum\limits_j^{N_{\beta}}\delta(r -r_{ij}^{(\beta)} ).
\end{equation}

For a completely sorted mixture, $g_{\beta \beta}(r)$ should exhibit an envelope that falls off exponentially, with a length scale $\xi$ that corresponds to the average cluster radius. In the HLT mixtures, where a single droplet forms, we see such a structure, as shown by the dashed black line in Fig.~\ref{fig:shapeG}d.  We extract a length scale of $\xi_{HLT} =4.5\pm 1.2$, which is very similar to the steady state average cluster radius shown in Fig.~\ref{fig:shapeG}c. For comparison, we measure $\xi$ for all shape bidisperse mixtures and compare it to $\xi_{HLT}$ by computing $\xi/ \xi_{HLT}$ (see inset to Fig.~\ref{fig:shapeG}d). We find this ratio to be quite small, consistent with previous results.

\noindent\textbf{Binary mixtures with two target areas.}
After studying the impact of shape disparity in cell sorting, we next study the effect of dispersity in area. The mixture is now comprised of equal numbers of cells with $A_{0,1}<A_{0,2}$, where we take $\sqrt{A_{0,1}}$ as the unit of length. Both types have the same $s_0$, or $s_{0,1}=s_{0,2}$. 
We have taken care to ensure that our area bidisperse mixture are also in a fluid region of the phase diagram (Fig.~\ref{fig:highKa}b) by checking that $D_{eff}>0.01$. For the results shown here, the shape index is fixed at $s_0=3.85$ to mimic fluid-like cells. We define the ratio of the preferred areas as $\alpha=A_{0,2}/A_{0,1}$.

\begin{figure}[!t]
\centering
\includegraphics[width=0.99\textwidth]{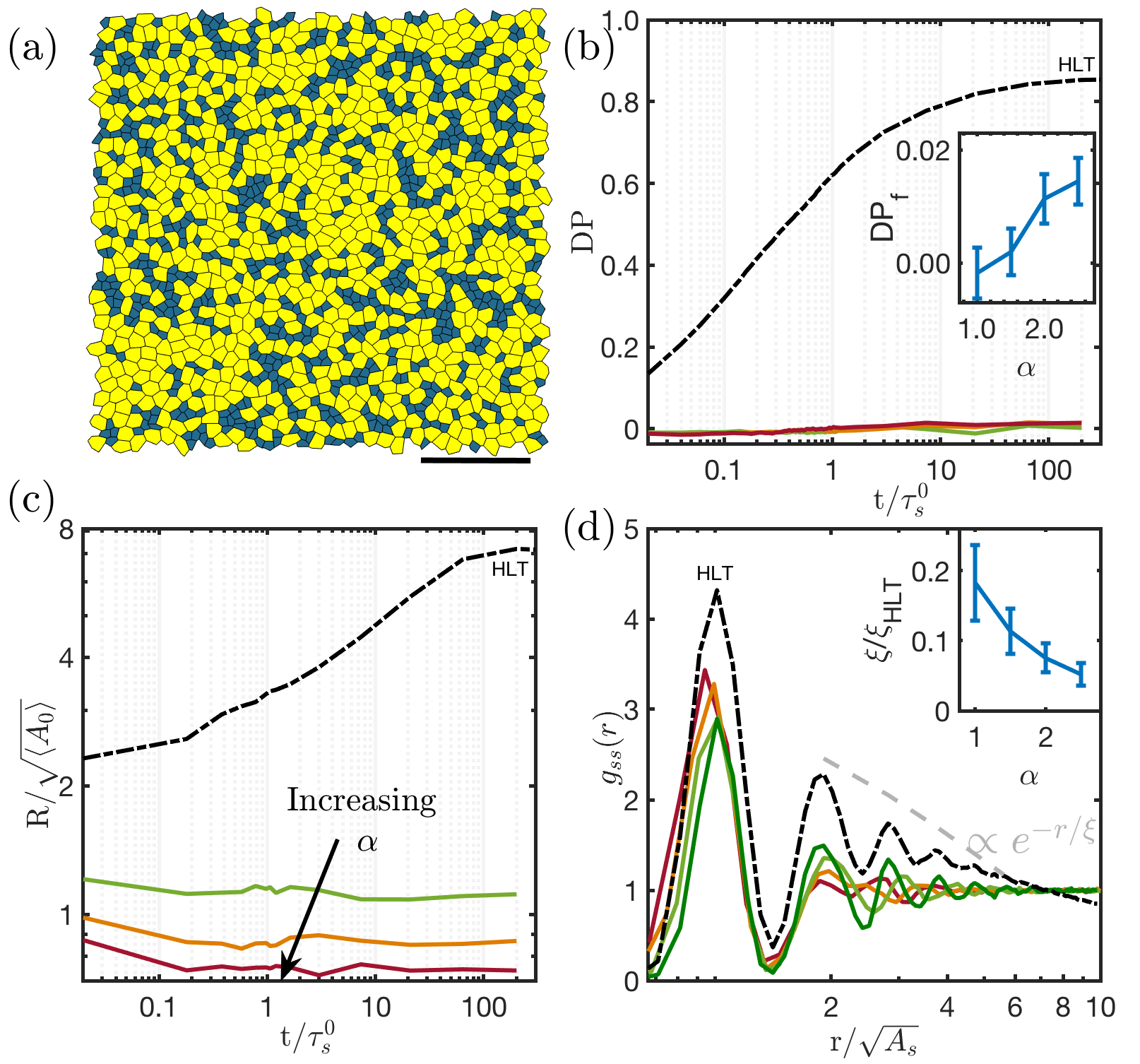}
\caption{ {\it Area bidisperse fluid mixtures.} (a) Snapshot of a $s_0=3.85$, $\alpha=2.5$, $N=1600$ mixture. Scale bar denotes 10 units. Yellow is used for larger cells ($A_0=1.43$) and blue for smaller ones ($A_0=0.57$). (b)-(d) Various quantifications of demixing in area bidisperse mixtures (curves colored from green to red in increasing order of size disparity i.e $\alpha=1.0,1.5,2.0,2.5$) are compared to a mixture with an extra heterotypic line tension of value 0.1, $s_0=3.97$ and $A_0=1.0$ (black dashed curve). (b) Demixing Parameter versus $\log(time)$. The final value ($DP_f$) as a function of $\alpha$ is shown in the inset. (c) Average cluster radius (R) versus time. (d) Pair correlation function of small-$A_0$ cells $(g_{ss})$ versus radial distance in units of the smallest lengthscale  for $t=200\,\tau_s^{0}$. The dashed grey line shows an exponential decay. The inset shows the decay lengthscale ($\xi$) in terms of the maximum possible lengthscale ($\xi_{HLT}$) with increasing disparity $\Delta$. Simulation details provided in Table S2.
}
\label{fig:areaR}
\end{figure}

Visual inspection of a snapshot from a simulation of an area bidisperse mixture with high $\alpha=2.5$ at long times demonstrates that observing cluster formation by eye is difficult, particularly given the disparity in area fraction between the two cell types (see Fig.~\ref{fig:areaR}). The DP has been measured and is smaller than those found in shape bidisperse mixtures (Fig.~\ref{fig:shapeG}b). Since the large-$A_0$ cells occupy more than half of the total area, we perform our cluster analysis on cells with $A_{0,1}$. As shown in Fig.~\ref{fig:areaR}c, the final clusters have an average cluster radius that is typically less than two cell diameters and becomes smaller as $\alpha$ increases. Similarly, Fig.~\ref{fig:areaR}d illustrates that $g_{ss}(r)$ also shows no sign of bulk demixing, with a structural length scale that is always less than $0.2*\xi_{HLT}$, and decreases with decreasing $\alpha$, as seen in the inset to Fig.~\ref{fig:areaR}d.

\noindent\textbf{Zero-temperature energy configurations.}
Our finite-temperature simulations suggest that large-scale sorting is not preferred in these mixtures. To understand this, we study an ensemble of energy minimized states. {\it If the mixed state has a lower energy at zero temperature, then we expect that energetics cannot drive demixing at finite temperature.} Therefore, we compare the energy of two initial states of $N=400$ cells: a sorted system where all of the cells with cell centers in the left half of the box are labeled type 1, and the remainder are labeled type 2, and a mixed system where cell types are randomly assigned. We use FIRE minimization~\cite{FIRE} to identify the nearest local energy minimum for $250$ realizations in each of the two scenarios.

\begin{figure}[h!]
\centering
\includegraphics[width=0.99\textwidth]{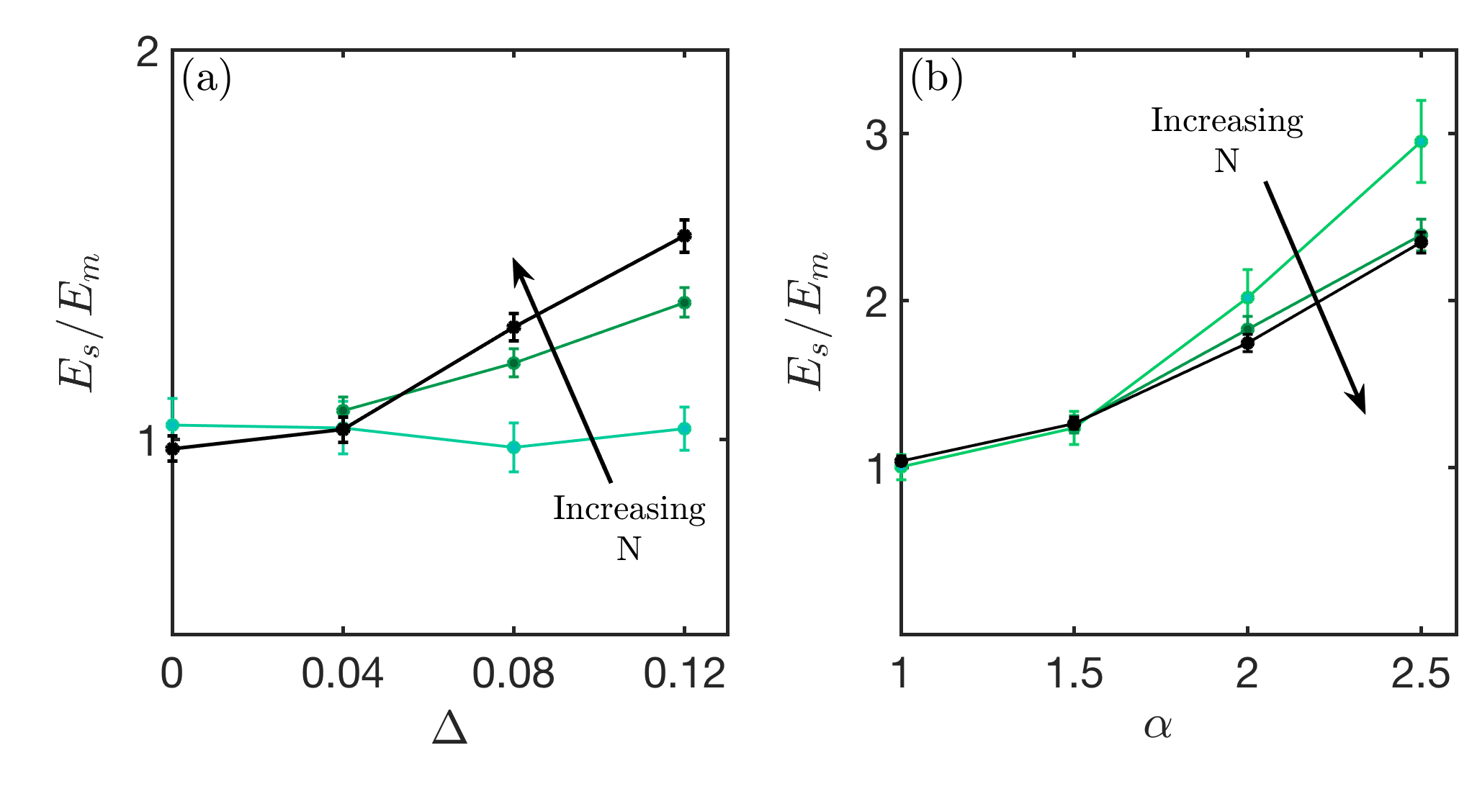}
\caption{{\it Minimal energy configurations.} Systems with $N=100,400,900$ cells (green to black) are energy minimized using the FIRE algorithm to get the total energy of the configurations- mixed($E_m$) and sorted($E_s$), for a $<s_{av}>=3.85$, $K_a=100$ with increasing disparity. The ratio $E_s/E_m$ versus $\Delta$. (b) The ratio $E_s/E_m$ is plotted versus $\alpha$ for $s_0=3.85$ and $K_a=1$. Simulation details provided in Table S3.}
\label{fig:cost}
\end{figure}

Figure~\ref{fig:cost}a shows the ratio between the energy of states with sorted initial conditions ($E_s$) and mixed initial conditions ($E_m$) in the case where type 1 and 2 cells have different shape parameters. At larger system sizes, there is a clear trend that the sorted states typically possess higher energies than the mixed states, so that the ratio rises above unity as the shape disparity $\Delta$ increases.  This indicates that there are no energetic forces driving the demixing in larger systems. We have also quantified the effective interfacial line tension (Fig.~\ref{fig:shapeTension}, Fig.~\ref{fig:areaTension}) using a method developed previously by some of us ~\cite{Yang12663}. We find that there is no emergent line tension in any of these mixtures, which is consistent with our energy calculations. In Fig.~\ref{fig:cost}b, which shows the ratio of energies between sorted and mixed states for cells with area dispersity, the trend is even clearer. Again, sorted states have significantly higher energy compared to mixed states as the area dispersity $\alpha$ increases.

\noindent\textbf{Zero-temperature T1 energy barriers.}
Although the zero temperature energy calculations above help us understand the lack of macroscopic demixing in mixtures with no heterotypic interfacial tension, they do not explain the small-scale demixing seen, for example, in Fig.~\ref{fig:shapeG}b. Since both cell types are subject to the same geometrical and topological constraints and rearrange via T1 transitions, we now turn to an energetic analysis of T1 transitions for the bidisperse system.

\begin{figure}[!t] 
\centering
\includegraphics[width=0.99\textwidth]{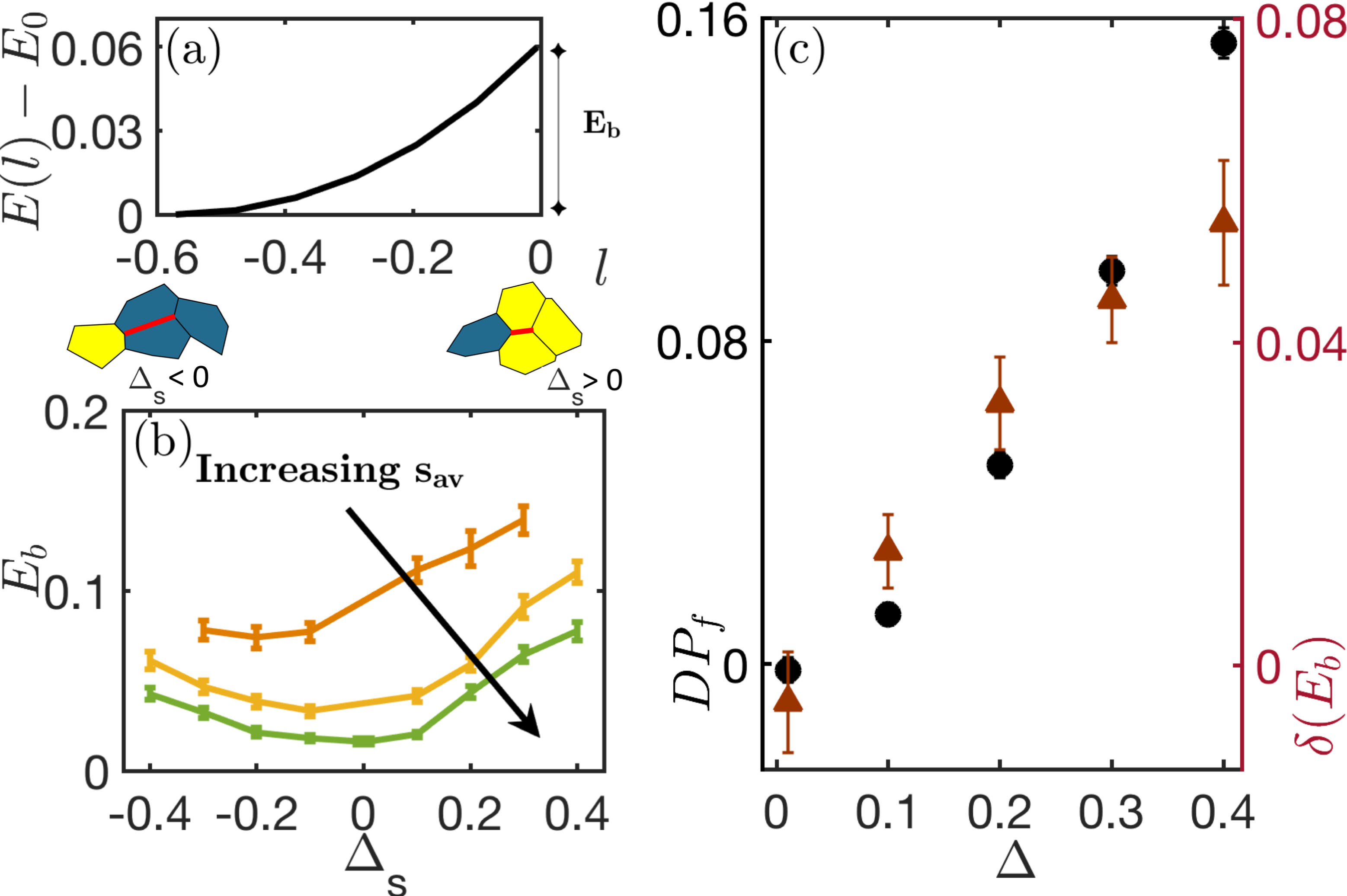}
  \caption{{\it Differential energy barriers in shape bidisperse fluid mixtures.} (a) Energy $E(l)$ relative to $E_0$ is plotted against T1 edgelength $l$ for a typical shape bi-disperse T1 pair ($\Delta=0.4, s_{av}=3.85$). (b) Energy Barrier $E_b$ is plotted against signed disparity in shape $\Delta_s$. Positive and negative $\Delta_s$ values imply stiffer cluster in yellow and floppier cluster in blue respectively, in the 4-cell diagrams show above. Each solid curve represents the barrier for a heterotypic cell to get out of the cluster for $s_{av}= 3.79,3.85,3.88)$ (from solid-like (orange) to liquid-like (green) (c) Correlation plot for $s_{av}=3.85$ between Differential Energy Barriers on the right y-axis $\delta(E_b)$ (in maroon triangles) and demixing relative to mixed scenario $DP_f$ on the left y-axis (in black discs). Shape difference $\Delta$ is plotted on x-axis. Simulation details provided in Table S4.}
\label{fig:corrSp}
\end{figure}

Specifically, we study the statistics of energy barriers in bidisperse systems, where there are nine types of T1 transitions possible.  While we present data in the supplemental information for symmetric cases where two of the cells are of type 1 and two of are type 2, we focus here on asymmetric systems where 3 of the cells are of one type and one is of another type. As illustrated by the 4-cell cluster diagrams in Fig.~\ref{fig:corrSp}, such 3:1 arrangements naturally represent the cost of one cell type invading an interface composed of cells of a distinct type, which determines the dynamic stability of such an interface. 

Similarly to previous work~\cite{Bi2015g,Bi2016a}, we compute the T1 energy barrier height by measuring the global tissue energy as we force a single edge in our bidisperse simulation to shrink to zero length while minimizing the energy and allowing the other degrees of freedom to relax, as shown in Fig.~\ref{fig:corrSp}a. The energy barrier $E_b$ 
we report in Fig.~\ref{fig:corrSp}b is the difference between the final energy $E(l=0)$ at the 4-fold vertex and the initial energy $E_0$, or $E_b=E(l=0)-E_0$, averaged over $250$ edges with the same topology in small simulated tissues with $N=80$ cells.

Figure~\ref{fig:corrSp}a illustrates a particular type of (3:1) T1 energy profile where a single cell with shape parameter $s_0^{1}$ invades a 3-cell cluster formed by cells with $s_0^{cluster}$. We define a {\emph signed} shape disparity $\Delta_{sign}=s_0^{1}-s_0^{cluster}$ to distinguish it from a T1 with cell types swapped. Negative $\Delta_{sign}$ indicates that a more stiff cell is invading a cluster of floppy cells. We have checked that energy barriers are statistically identical for cells entering or exiting a cluster. Because cells are as likely to leave a cluster as to enter it, this suggests that clusters of a given cell type will not grow or shrink over long-time or length-scales.

Figure~\ref{fig:corrSp}b highlights that the energy barriers associated with these (3:1) transitions systematically increase as the magnitude of the shape dispersity $\Delta_{sign}$ increases. In other words, it becomes energetically more difficult for a single cell to invade or leave a cluster of a different cell type as the shape dispersity between the two types increases. Perhaps more importantly, it also shows that these energy barriers are not symmetric around zero; there is a systematic difference between a stiffer cell invading a floppier cluster and vice-versa, especially for lower values of $s_{av}$ as the system approaches the jamming transition. Stiffer clusters tend to be more difficult to break up than floppier clusters. To characterize this effect, we define the energy barrier disparity between invading stiffer and floppier clusters as $\delta E_b (\Delta)= E_b (\Delta)- E_b (-\Delta)$.

To test whether this mechanism might be relevant for the micro-demixing we observed in our finite-temperature simulations, we directly compare the demixing parameter associated with the final, steady state in each simulation, $DP_f$ to the energy barrier disparity $\delta E_b$ as a function of shape dispersity $\Delta$, as shown in  Fig.~\ref{fig:corrSp}c.  This plot shows a quite strong correlation between the two quantities, suggesting that this mechanism is a very likely driver of micro-demixing. To further test this idea, we have increased the temperature for the $\Delta=0.2$ mixtures and found $DP$ to vanish at temperatures higher than the differential energy barrier, as shown in  Fig.~\ref{fig:DiffTemp}.

A similar analysis can be performed for area bidisperse mixtures as shown in Fig.~\ref{fig:corr}. An important difference from the shape bidisperse case is that while there is a clear connection between cell shape and tissue rheology (stiffer cells have a smaller $s_0$), there is no such connection between area and rheology.  Moreover, there is very little evidence for micro-demixing, and so we expect the signal to be much weaker.  Nevertheless, we can define a quantity  $\alpha_s=A_0^{cluster}/A_0^{1}$ that is less than unity if a larger cell is invading a cluster of smaller cells and greater than unity otherwise. Fig.~\ref{fig:corr}b suggests that large-cell clusters are more difficult to invade than small-cell clusters, although the differential energy barrier is quite a bit smaller than for the case of shape bidispersity. In particular, the case where $s_0=3.85$ is highlighted in Fig.~\ref{fig:corr}c showing a correlation between demixing and $\delta E_b$, although the amplitude of both effects is quite small. \\

\begin{figure}[t!] 
\centering
\includegraphics[width=0.9\textwidth]{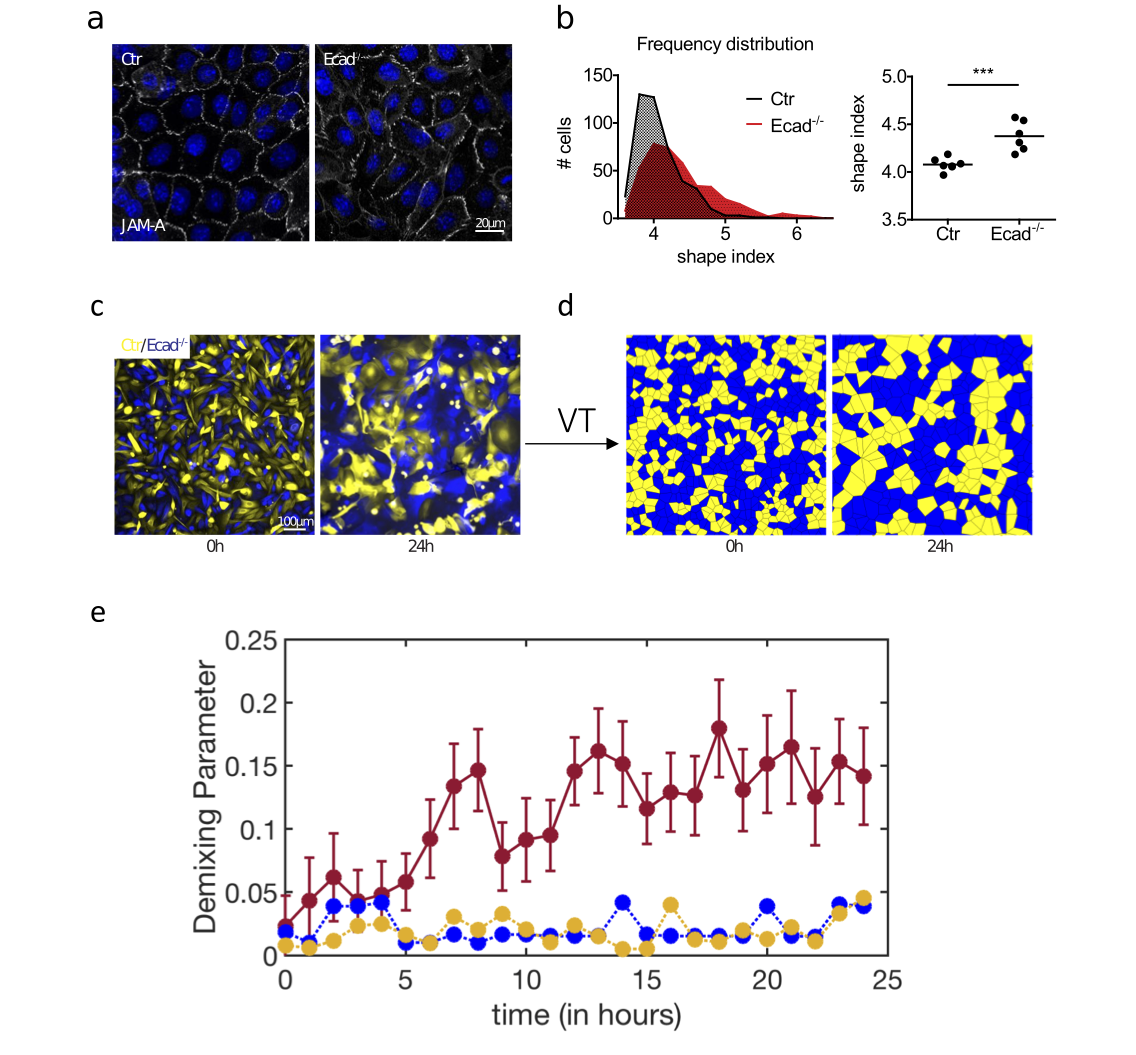}
  \caption{{\it Micro-demixing observed in keratinocyte co-cultures.} (a) Wild-type (Ctr) and E-cadherin knockout  (E-cad$^{-/-}$) celll monolayer mixtures with nuclei labelled using immunofluoroscence. (b) Histograms of cell shapes for Ctr and E-cad$^{-/-}$ cells are compared across seven and six different isolates respectively,  i.e. seven different mice. There is a clear difference in the shape index ($\Delta=0.31$) of both cell-types, with shape indices of $4.08 \pm 0.06$ for Ctr cells and $ 4.38\pm 0.14$ for Ecad$^{-/-}$ cells. (c) Both cell-types, Ctr in yellow and E-cad$^{-/-}$ in blue, start off initially mixed as shown the 0h (zero hours) snapshot. The co-culture evolves into a micro-segregated mixture, as shown in a 24h snapshot. (d) Voronoi tessellations (VT) of the same snapshots. (e) Solid maroon curve represents the time evolution of demixing parameter for the E-cad$^{-/-}$ cell-type in the mixture as a function of time and averaged over five different isolates. This result is compared against a control of demixing curves for some initially well-mixed regions of monolayers of either all Ctr cells or all E-cad$^{-/-}$ cells but with half the cells stained differently than the other half.  Well-mixed regions of the control cells are shown in yellow, while the well-mixed regions of the E-cad$^{-/-}$ cell-type is shown in blue. }
\label{fig:expt}
\end{figure}

\bigskip
\noindent\textbf{\large Experimental Results}\\
To test our modeling against experiments, we first study monolayers of primary keratinocytes (Ctr) in the presence of high calcium, whose presence initiates intercellular junction formation (see Fig. 6a). Under such conditions, the monolayer is confluent in the sense that there are essentially no gaps between cells. We also test whether the  confluent monolayer is fluid-like by measuring the displacement of cells over the course of 24 hours. While some number of neighbor exchanges indeed take place, and while the integrated displacement of the cells is several times a typical cell length, we find that the mean square displacement of these cells is typically of the order of a single cell size (Fig. S10a). This places some limitations on the scale of demixing that is expected, but we nevertheless see a level of micro-demixing that is comparable to what we observe in our simulations (Fig.~\ref{fig:corr}e).

Since the shape index is an important parameter in our theory, we measure this quantity for Ctr cells in the monolayer and obtain an average shape index of $4.08\pm0.06$. The full histogram is plotted in Fig. 6b. Confluent monolayers of primary keratinocytes but with E-cadherin knocked-out, or E-cad$^{-/-}$ cells, again in the presence of high calcium, are then studied to check for confluency (Fig. 6a) and fluidity (Fig. S10a). We then measure an average shape index of $4.38\pm0.14$ for the E-cad$^{-/-}$ cells (Fig. 6b). A T-test reveals that the difference in the two shape index histograms is statistically significant with a P-value of 0.0008. The difference in the shape index corresponds to $\Delta=0.31$.  Since we explore both differential adhesion and differential size, we also measure the areas of each cell type in the monolayer and found no statistically significant area difference. See Fig. S10(b). In other words, the monolayer mixture tests differential adhesion, as opposed to both differential adhesion and bidisperse areas. 

Next, we study the monolayer of approximately a 50:50 Ctr/E-cad$^{-/-}$ mixture in the presence of high calcium over the course of 24 hours.  Again, a major complication in our comparison is that while both types of cells in the mixture exert active forces on their environment the typical displacements over this time frame are small. Nevertheless, after constructing a Voronoi tessellation for snapshots of the monolayer taken every hour, we measure the demixing parameter (DP) for each cell type, accounting for the fact that the mixture is not precisely a 50:50 mixture (see Fig. S10c). In doing so, we measure number of neighbors for each cell type and subtract off the corresponding number fraction. Figure 6e shows the DP parameter as a function of time for the E-cad$^{-/-}$ cells in the mixture. This parameter increases from zero (within one standard deviation) and appears to saturate after approximately 19 hours to around 0.15, albeit with some fluctuations. Such values of the DP parameter are consistent with our computational observations of micro-demixing in the differential adhesion case. 

We argue that these results suggest the small-scale demixing is a consequence of large differences in differential adhesion. To rule out this being a consequence of the natural variability in adhesion even within one cell type, we also measure the demixing parameter in monolayers of just Ctr cells and of just E-cad$^{-/-}$ cells by staining half of the cells with one type of stain and the remaining half with a second type of stain (checking for an artificial ``demixing'' due to variability in these two monotypic monolayers). We find that the demixing parameter does not increase over time on average for either the Crt cells or the E-cad$^{-/-}$ cells when looking at initially well-mixed regions (Fig. 6e) or the entire monolayer (Fig. S11a).  In addition to the DP, we also compute the pair-correlation function for the experimental system and find that while it contains less structure than the simulations, it also exhibits a correlation length over several cell diameters (see Figs. S11b and c).

While the qualitative and semi-quantitative comparison between the degrees of micro-demixing observed in experiments and vertex model simulations is promising, we must also acknowledge that there are several differences between the two settings.   Taking into account such differences and determining how the micro-demixing is potentially affected is a future avenue for investigation. For instance, our computations so far focus on 50:50 mixtures and do not take into account the potentially persistent motion of cells. Another difference is the apparent timescale over which the micro-demixing occurs from various initial conditions: in the experiments the demixing seems to occur during a time in which the cells move not much more than a typical cell size; in contrast our simulations require many $\tau_\alpha$ to reach comparable levels of demixing from a random initial configuration. Additionally, while we expect the differential energy barriers to remain at least for some range of persistence, the value of the DP parameter will depend on that just as the steady state value of the DP parameter depends on temperature in our over-damped Brownian simulations. Further differences may include the effect of differential motility, differential mechanical stiffnesses of the cells, and differential cell division and death (which can itself affect the diffusivity of cells~\cite{cellbirth}). Therefore, to more rigorously test the computations against the experiments, future experimental work with detailed cell tracking within the monolayer and the prevention of cell birth with the introduction of mitomycin, as well as additional computational work, needs be implemented.

\bigskip
\noindent\textbf{\large Discussion}\\
Using Brownian vertex model simulations, we show that two-dimensional mixtures, bidisperse in preferred shape and in preferred area, have robust fluid-phase mixing at large scales in the absence of an explicit heterotypic line tension distinguishing between the two cell types. Energy minimization at zero temperature further supports this finding: mixed systems have lower energy than sorted ones, so that bidispersity is not sufficient to energetically stabilize an interface between the two fluids. For shape bidisperse mixtures, we find that, in spite of having solid-like cells making up half the mixture, the mixtures are still able to fluidize in some parameter regimes of the vertex model. Furthermore, although this large scale mixing occurs, we find persistent and equally robust micro-demixing in shape bidisperse mixtures, where clustering of the same cell type over sub-system-spanning lengthscales is observed.

To understand micro-demixing in shape bidisperse mixtures, we establish a correlation between micro-demixing and zero-temperature differential energy barriers for neighbor exchanges (T1 transitions) between four cells at the heterotypic boundaries. Specifically, we find that the energy barriers for a fluid cell type to ``invade'' a cluster of stiff cells is typically higher than for a stiff cell to ``invade'' a cluster of fluid cells. This difference in energy barriers creates a bias towards the small-scale clustering of stiffer cells. For area bidisperse systems, the differential energy barriers for neighbor exchanges are smaller than for the shape bidisperse case, and we find a negligible amount of micro-demixing. Our differential energy barrier calculations at zero temperature also yields a prediction for the temperature above which the micro-demixing does not occur---a prediction that has indeed been verified in our simulations.  

The computational observation of robust mixing on large scales for both types of mixtures may be surprising, given that the shape-based interaction distinguishes between the two cell types just as changing the strength of the distance-dependent interaction between two particles of different types in thermal Lennard-Jones mixtures. In the particulate case, there is either large-scale demixing or no demixing (depending on the miscibility), while in the cellular case, there can be micro-demixing. This suggests that vertex models may be more relevant for characterizing cell sorting in dense cellular mixtures than other coarse-grained modeling approaches. 

What about comparisons with athermal particle systems? Athermal two-dimensional bidisperse particulate mixtures of different size discs with purely repulsive forces, such as models for granular particles with no (or little) friction, are not expected to sort at small size disparities~\cite{Lopez-Sanchez2013a}. Only as the size dispersity increases does sorting occur due to entropic depletion forces~\cite{Melby2007}. Entropic depletion forces do not apply to a confluent tessellation in which the packing fraction is fixed at unity, though may to some extent apply to Voronoi models. Depletion forces also drive demixing in vertical vibrated shape bidisperse mixtures of rods and spheres~\cite{Rodriguez-Linan2016}. Interestingly, size bidisperse mixtures of active particles can sort in the absence of any attractive forces~\cite{Yang2014a}. The sorting here is due to an asymmetry in the energy barrier between one smaller particle passing through two larger particles as compared to one larger particle passing through two smaller ones. Given the above analogy, a vertex model fluid mixture perhaps has more in common with an active, disordered binary packing than with a thermal fluid mixture with differential adhesion. 

How different is the vertex model examined here applied to cells in comparison to the vertex models applied to foams? The robust mixing observed in area bidisperse systems at zero temperature is indeed  counter-intuitive when compared with area bidisperse foams in ordered hexagonal states~\cite{Graner2002}. In this case, the system demixes at zero temperature given an additional perturbative energetic cost to an interface between cells of slightly different areas. Only for large applied shear strains do area bidisperse foams mix~\cite{Cox2006}. Understanding differences between foam and vertex models is therefore an interesting area for future study. Foam models lack the $P^2$ contribution to the energy functional (Eq.\ref{eq:energy}) and this restricts the fluid-like phase space accessible to such models, perhaps contributing to differences between them.

To determine whether or not our micro-demixing prediction is directly relevant for biology, we conduct experiments with cellular monolayers consisting of both wild-type keratinocytes and E-cadherin-knock-out keratinocytes. Such mixtures allow us to study differential adhesion and its effect on cell sorting in the absence of heterotypic tensions. We find evidence for micro-demixing with a saturated demixing parameter that agrees with our prediction to within one standard deviation of the experimental value. Moreover, we do not observe large-scale demixing over the time scale of the experiment. Over longer time scales, the monolayers gradually become multi-layered, an effect we have yet to incorporate into our computational modeling of confluent cellular mixtures. In addition to cell persistence, differential motility, and cell birth/death, we also have yet to explore the effects of small lapses in confluency, which could arise given the lack of E-cadherin in the modified cell type. Such exploration will allow for even more rigorous quantitative comparison between the modeling and the experiments.

Our computational and experimental results bring an understanding to earlier work demonstrating that sorting at embryonic boundaries requires high heterotypic interfacial tension~\cite{Canty2017b}. Given our T1 energy barrier analysis encoding both the topological and geometrical constraints of confluent packings, we now understand why these mixtures robustly mix. This robustness suggests that despite some difference in shape and size, progenitor cells can readily mix throughout the embyro. To demix (or sort), progenitor cells have developed biochemical means of recognizing whether neighboring cells are of the same type or a different types. And while a small amount of heterotypic line tension can generate stable interfaces~\cite{Sussman2018b} in the absence of fluctuations, correlated fluctuations may be able to overcome such barriers. Our analysis gives a new way to understand bulk behavior based on cellular rearrangements in such confluent mixtures. In other words, based on the analysis of T1 energy barriers between different cell types, experimentalists can predict whether or not different cell types will mix or not mix in the bulk.

Finally, the micro-demixing effect observed both in our computations and in our experiments could be utilized in biology to create more subtle patterning.  For instance, when randomly tagging a tessellation half with one cell type (and half with another cell type), one of the cell types percolates through the system~\cite{Bollobas2006}. However, if there is now some spatial correlation in the tagging introduced even at the small scale, such that the tagging of one cell type is positively correlated with tagging a neighboring cell of the same type, then the percolation transition point can be altered, transitioning from a tenuous spanning structure to one that is more robust and more able to respond to changes in the environment.

\noindent\textbf{\large Methods and Materials}    \\
\noindent\textbf{Simulation details:} We simulate a vertex model where the degrees of freedom evolve according to over-damped Brownian dynamics~\cite{Sussman2017a}. Specifically, each vertex $i$ located at coordinate $\bf{r}$ experiences a Brownian force  $\mathbf{F}^B$ with $\mathbf{F}_{i}^{B}=${\boldmath$\xi$}$_i$\label{eq:dynbm}, where {\boldmath$\xi$}$_i$ is white noise with zero mean and $\langle\xi_{\gamma i}(t)\xi_{\lambda k}(t')\rangle=2T\delta_{\gamma\lambda}\delta_{ik}\delta(t-t')$ with $\gamma$ and $\lambda$ denoting spatial components and in units of $k_B$ equal to unity. In epithelial layers, we expect that fluctuations are driven by active cytoskeletal components, and hence the $T$ is an effective temperature that represents the magnitude of this activity (Ref.~\cite{Gunst2015}). The equation of motion for a single vertex, therefore, takes the form
\begin{equation}
\dot{\mathbf{r}}_{i}=\mu \mathbf{F}_{i} +\mu \mathbf{F}_{i}^B,\label{eq:dyn} 
\end{equation}
with $\mathbf{F}_{i} =-${\boldmath$\nabla$}$_i E$, where E is the total energy as defined in the Model section. The force $\mathbf{F}_i$ is a non-local effective mechanical force experienced by the $i$th vertex of the $j^{th}$ cell and hence represents the cell-cell interactions. In the absence of mechanical interactions, an isolated cell performs a random walk with a long time effective diffusion rate of $T/\mu$. Unless otherwise specified, $\mu=1$.  Finally, the Euler-Murayama integration method is used to update a discretized version of the equations (one for each vertex). One simulation unit time is referred to as $\tau$.  For a system with no dispersity, a cell with a shape parameter $s_0$ of $3.85$, typically requires $1000\tau$ to move its own length, i.e. $\tau_s^{0}=1000\tau$. We typically simulate up to times several hundred times greater than $\tau_s^{0}$.  We also note that other models with directed cell motility are possible, including an Active Vertex Model~\cite{Czajkowski2018} and the SPV model where cells are self-propelled due to an active force with persistence~\cite{Bi2016a,Barton2017}. 

In vertex models, one needs to take care of cellular rearrangements explicitly~\cite{Farhadifar2007b, Bi2016a}. In the absence of cell division or death, such rearrangements correspond to T1 transitions in which one edge shrinks to zero length and two new cells are connected via a new growing edge.  In simulations, if an edge length falls below some threshold length $l_c$, then we rotate the edge by $\pi/2$ and reconnect the topology of the surrounding cells to generate a local neighbor-exchange. Unless otherwise specified, $l_c$ is set to 0.04. The noise is controlled by temperature (T) which is set to 0.01. 

Past work has demonstrated that the mechanical properties of vertex models depend sensitively on the shape parameter $s_0$ and temperature $T$. Specifically, these models exhibit rigidity~\cite{Bi2015g} or glass~\cite{Sussman2018c} transitions where the system transitions from more solid-like to more fluid-like. At $T=0$, the 2D vertex model exhibits a rigidity transition as a function of cell shape parameterized by $s_0$. Above a critical value of target shape index $s_0^* \sim 3.81$, cells are able to move past each other with very small energy cost and below which they cannot. To understand this transition, one analyzes the energetics of how cells move past each other via T1 transitions. A minimal four cell calculation with fixed unit area hexagonal cells revealed that if the two cells that would no longer share an edge after the edge swap formed regular pentagons, then the energy barrier for the formation of a four-vertex vanishes, suggesting that pentagon shape formation is a geometrically compatible transition pathway for three-fold coordinated lattices. Interestingly, the shape parameter for a regular pentagon is $s_0=2 \sqrt{5} (5-2\sqrt{5})^{1/4}\approx 3.812\approx s_0^*$.  In the presence of activity or temperature, vestiges of this zero-temperature rigidity transition have been found in a glassy transition between fluid-like and more solid-like behavior in an active Self-Propelled Voronoi (SPV) model~\cite{Bi2016a} and a Brownian Voronoi model~\cite{Sussman2018c}.

Given the complex phase behavior of such vertex models, we want to ensure that the mixtures are fluid-like.  To do so, we first measure the Mean-Squared Displacement (MSD). To account for global tissue motion possible in these types of models~\cite{Giavazzi2018}, we define the displacement of each cell in a time window $t$, $\mathbf{x}(t)$, as the distance the cell traveled in time $t$ minus the total displacement of the entire system of cells over that same time interval.  Then the MSD is defined as
\begin{equation}
\label{eq:msdDef}
MSD(\Delta t)  \equiv  \langle (\mathbf{x}(t + \Delta t)-\mathbf{x}(t))^2 \rangle,
\end{equation}
where $\langle \cdot \rangle$ denotes an average over all cells in the tissue and all times $t$. The self-diffusivity $D_{s}$, is defined by assuming the long-time behavior of the system is diffusive,
\begin{equation}
 \label{eq:msdDs}
 D_s=\lim_{t\rightarrow\infty} \frac{MSD(t)}{4t}.
\end{equation}
To understand whether cells are being constrained by their neighbors, we compare $D_{s}$ to the \textit{bare}, or non-interacting, diffusion constant $D_0$. For a non-interacting Brownian particle at temperature T with mobility $\mu$, the Fluctuation-Dissipation theorem states that $D_0=\mu k_B T$, where $k_B$ is Boltzmann constant. We set $\mu k_B$ to unity.  The effective diffusivity is given by
\begin{equation}
 \label{eq:Def}
D_{eff}= \frac{D_s}{D_0 }.
\end{equation}
 Systems with small $D_{eff}$ are more solid-like, while systems with large $D_{eff}$ are more fluid-like.  In practice, we use a threshold of $0.01$ to distinguish between these different behaviors, in line with previous work~\cite{Bi2016a}.\\
 
 Self-diffusivity time $\tau_s$ is defined as the time taken by a cell to move its own length. For a 2D system, one can use Eq:\ref{eq:msdDs} to compute $\tau_s=1/4D_s$. Dispersity in the system can affect this average motion. Hence we convey time in units of $\tau_s^{0}$ which we define as the self-diffusivity of cell, with $s_0$ of $3.85$, in absence of any dispersity. To study micro-demixing, we run simulations  that are $200 \,\tau_s^{0}$ long i.e. long enough for cells to explore the entire system multiple times. 

\noindent\textbf{Experimental details} \\
\noindent{\it Isolation and culture of primary keratinocytes}\\
Primary keratinocytes isolated from newborn mice were cultured in DMEM/HAM’s F12 (FAD) medium with low Ca$^{2+}$ (50 μM) (Biochrom) supplemented with 10 \% FCS (chelated), penicillin (100 U ml$^{-1}$), streptomycin (100 $\mu$g ml$^{-1}$, Biochrom A2212), adenine ($1.8\times 10^{−4}$ M, SIGMA A3159), L-glutamine (2mM, Biochrom K0282), hydrocortisone (0.5 $\mu$g ml$^{-1}$, Sigma H4001), EGF (10 ng ml$^{-1}$, Sigma E9644), cholera enterotoxin (10$^{−10}$ M, Sigma C-8052), insulin (5 $\mu$g ml$^{-1}$, Sigma I1882), and ascorbic acid (0.05 mg ml$^{-1}$, Sigma A4034). For keratinocyte isolation newborn mice were sacrificed by decapitation and incubated in 50\% Betaisodona/PBS for 30 minutes at 4$^o$C, 1 minute PBS, 1 minute 70\% EtOH,  1 minute PBS and 1 minute antibiotic/antimycotic solution. Tail and legs were removed and complete skin incubated in 2 ml Dispase (5mg ml$^{-1}$)/FAD solution. After incubation over night at 4$^°$C, skin was transferred onto 500 $\mu$l FAD medium on a 6 cm dish and epidermis was separated from the dermis as a sheet. Epidermis was transferred dermal side down onto 500 $mu$l of TrypLE (ThermoFisher Scientific) and incubated for 20 minutes at RT. Keratinocytes were washed out of the epidermal sheet using 3 ml of 10\%FCS/PBS. After centrifugation keratinocytes were resuspended in FAD medium and seeded onto Collagen type-1 (0.04mg ml$^{-1}$) (Biochrom, L7213) coated cell culture plates. Primary murine keratinocytes were kept at 32$^°$C and 5\% CO$_2$. To induce classical cadherin dependent junction formation, cells were switched to high Ca$^{2+}$ medium (1,5-1,8 mM). Cultured cells were regularly monitored for mycoplasma contamination and discarded in case of positive results. E-cadherin-/- keratinocytes were isolated from E-cadherin epidermal knockout mice as described previously~\cite{Rubsam}.\\

\noindent{\it Keratinocyte labeling and demixing assay}\\
Keratinocytes were resuspended according to 500000 cells/ml in 1 ml dyeing solution (medium with 10 $\mu$M CellTracker™ Green or CellTracker$^{TM}$ Orange (ThermoFisher \#C7025 or \#C34551 respectively, stock 10 mM in DMSO)). Cells were incubated for 20 minutes at 32$^o$C and pelleted at 850 rpm for 5 minutes. Cells were resuspended in  1ml medium and incubated for another 30 minutes at 32$^o$C. Eventually, cells were pelleted again and resuspended in 1 ml and green and orange cells of equal numbers were mixed and plated in low Ca$^{2+}$ FAD medium. Cell numbers were chosen to achieve confluency immediately after attachement and spreading. Twenty-four hours after plating medium was changed either low or high Ca$^{2+}$ FAD medium to induced cell-cell junction formation and demixing. Prior to live cell imaging, cells were incubated in FAD medium containing Hoechst dye to label nuclei for 1 hour. During live cell imaging cells were kept under controlled temperature (37$^o$C) and CO$_2$ (5\%) and imaged every hour.\\

\noindent{\it Immunofluorescence of keratinocytes in vitro (cell shape)}\\ 
For immunofluorescence staining of keratinocytes, cells were seeded on collagen coated glass cover slips in a 24 well plate and switched to high Ca$^{2+}$ medium at confluency for 2 hours. Cells were fixed using 4 \%PFA for 10 minutes at RT, washed three times for 5 minutes using PBS, permeabilized using 0. 5\%TritonX100/PBS and blocked using 5\% NGS/1\%BSA/PBS for 1 hour at room temperature. Primary antibodies were diluted as indicated in the antibody section in Background Reducing Antibody Diluent Solution (DAKO). Cover slips were placed growth surface down onto a 50$\mu$l drop of staining solution on parafilm in a humidified chamber and incubated over night at 4$^o$C. Cover slips were transferred back into the 24 well plate and washed with PBS three times for 10 minutes. Secondary antibodies and DAPI (4’,6-Diamidin-2-phenylindol, Sigma) were diluted 1:500 in PBS and cover slips were incubated for 1 hour at RT. Secondary antibodies were washed off via three wash steps using PBS for 10 minutes. Cover slips were mounted using Gelvatol (Calbiochem).\\

\noindent{\it Antibodies and inhibitors} \\
Primary antibodies used in this study: 
rat monoclonal against ZO-1 (hybridoma supernatant~\cite{Stevenson}, clone R26.4C); rat monoclonal against JAM-A (1:300, clone H2O2-106-7-4, kind gift from Sandra Iden). Secondary antibodies were species-specific antibodies conjugated with either AlexaFluor 488, 594 or 647, used at a dilution of 1:500 for immunofluorescence (Molecular Probes, Life Technologies).\\

\noindent{\it Microscopy}\\
Confocal images were obtained with a Leica TCS SP8, equipped with a white light laser and gateable hybrid detectors (HyDs). Objectives used with this microscope: PlanApo 63x, 1.4 NA. Epifluorescence images were obtained with a Leica DMI6000. Objectives used with this microscope: PlanApo 63x, 1.4 NA; PlanApo 20x, 0.75 NA.\\

\noindent{\it Image processing and analysis}\\
Quantification of cell shapes: Keratinocyte cell shapes were analyzed 2 hours after Ca$^{2+}$ switch and induction of cell-cell junction formation. Cell-cell boundaries were labeled by staining for one of two early cell-cell junction markers ZO-1 or JAM-A. Images were analyzed using Fiji~\cite{Schindelin}. Cell-cell boundaries were delineated manually using the polygon tool and perimeter and area were measured to calculate the shape index as described above.

Keratinocyte demixing: Leica Imaging Files (LIF) were analyzed using cell profiler 3.1.9~\cite{McQuin}. Nuclei of cells were identified and nuclei areas were used to measure green fluorescence to discriminate between green and red cells.\\

\noindent\textbf{Code availability:} 
The codes are programmed using open source cellGPU code available at \href{https://github.com/sussmanLab/cellGPU}{https://github.com/sussmanLab/cellGPU}.\\
\noindent\textbf{Data availability:} 
The data that support the findings of this study are available from the corresponding authors upon reasonable request. \\
\bibliography{2dmixtures}
\noindent\textbf{\large Acknowledgments} We would like to thank Matthias Merkel for useful discussions. This work was supported by NSF-POLS-1607416 (MCM, MLM, JMS) NSF-DMR-1352184 (MLM), Simons Foundation-454947 (MLM), Simons Foundation-342345 (MCM), SFB 829 A1 (CM), SPP1782 (CM), NI 1234/6 (CM). SP, DMS, MLM, and JMS also acknowledge support of the Syracuse University Soft and Living Matter Program and BioInspired Syracuse.\\
\noindent\textbf{\large Author contributions}: 
All authors conceived the study; P.S. conducted the simulations; M.R. and A.M. conducted experiments; P.S., D.M.S., A.F.M. and M. R. contributed to the analysis; all authors interpreted the results and wrote the paper.\\
\noindent\textbf{\large Additional information}: \\
\textbf{Competing interests:} The authors declare no competing financial or non-financial interests. 
\pagebreak

\section*{Supplemental Material}
\beginsupplement

\subsection*{Simulation Parameters}

Here we provide tables for the parameters used for each aspect of the simulations. For our dynamical simulations, the systems are equilibrated for time, $t_{eq}$, and subsequently run for a longer time. For our FIRE simulations, simulations typically run until the maximum force experienced by a vertex reduces below a threshold value of $10^{-13}$.

\begin{table}[h]
\begin{minipage}{.5\textwidth}

\centering
\caption{Shape Bi-disperse Dynamical Simulations}
\begin{tabular}{lc}
Parameters & values  \\
\hline
1. Ensembles & 50  \\
2. $s_{av}$ & 3.79- 3.91 \\
3. $\Delta$ & 0.0-0.4 \\
4. $t_{eq}$ & 1000 \\
5. $dt$ & 0.001 \\
6. $(K_a,K_p)$ & (100,1) \\
7. $T$ & 0.01 \\
8. $N$ & 400 \\
9. Total time & $2\times\,10^5 + t_{eq}$ \\
10. $l_c$ & 0.04 \\
11. HLT $(\gamma)\,\,\, for\,\,\, p_0=3.97$  & 0.1 \\
\hline
\end{tabular}
\label{tab:shape_dynamical}

\end{minipage}
\begin{minipage}{.5\textwidth}

\centering
\caption{Area Bi-disperse Dynamical Simulations}
\begin{tabular}{lc}
Parameters & values  \\
\hline
1. Ensembles & 50  \\
2. $s_{0}$ & 3.85 \\
3. $\alpha$ & 1.0-2.5 \\
4. $t_{eq}$ & 1000 \\
5. $dt$ & 0.01 \\
6. $(K_a,K_p)$ & (1,1) \\
7. $T$ & 0.01 \\
8. $N$ & 400 \\
9. Total time & $2\times\,10^5 + t_{eq}$ \\
10. $l_c$ & 0.04 \\
11. $\langle A_0 \rangle$ & 1 \\
\hline
\end{tabular}
\label{tab:area_dynamical}

\end{minipage}
\end{table}


\begin{table}[h]
\begin{minipage}{.45\textwidth}

\centering
\caption{FIRE minimization for $E_s/E_m$}
\begin{tabular}{lc}
Parameters & values  \\
\hline
1. Ensembles & 250  \\
2. $s_{av},s_0$ & 3.85 \\
3. $\alpha$ & 1.0-2.5 \\
4. $\Delta$ & 0-0.12 \\
5. $dt$ & 0.01 \\
6. $K_a$ & 1$(\alpha)$ \& 100 $(\Delta)$ \\
7. $T$ & 0.01 \\
8. $N$ & 100,400,900 \\
9. Maximum FIRE steps & $10^5 $ \\
10. $l_c$ & 0.04 \\
\hline
\end{tabular}
\label{tab:InterfaceCost}

\end{minipage} \qquad
\begin{minipage}{.45\textwidth}

\centering
\caption{T1 energy barriers}
\begin{tabular}{lc}
Parameters & values  \\
\hline
1. Ensembles & 250  \\
2. $s_{av}$ & 3.79-3.88 \\
3. $\alpha$ & 1.0-2.5 \\
4. $s_0$ & 3.82-3.88\\
5. $dt$ & 0.01 \\
6. $K_a$ & 1$(\alpha)$ \& 100 $(\Delta)$ \\
7. $T$ & 0.01 \\
8. $N$ & 80 \\
9. Maximum FIRE steps & $10^5 $ \\
10. $l_c$ & 0.04 \\
11. $\Delta$ & 0-0.12 \\
\hline
\end{tabular}
\label{tab:T1FIRE}

\end{minipage}
\end{table}

\subsection*{Effect of area stiffness on fluidity }
\begin{figure}[h] 
\centering
\includegraphics[width=\textwidth]{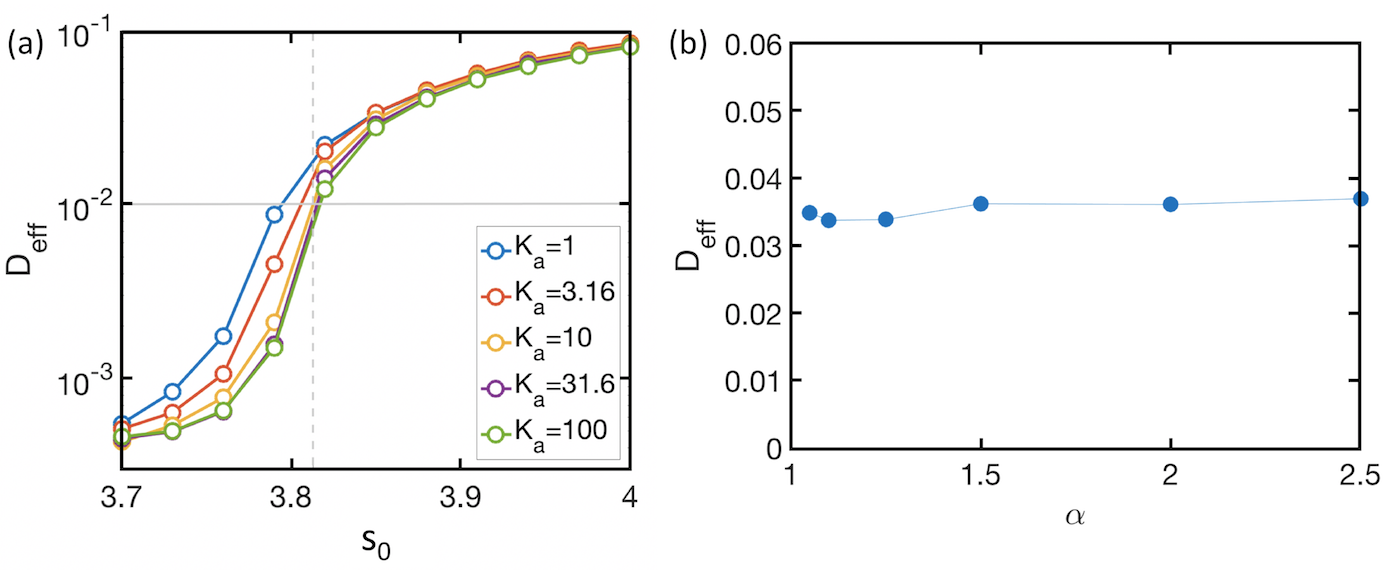}
  \caption{(a){\it Effective diffusivity ($D_{eff}$) with respect to target shape parameter $s_0$.} Different curves are for monodisperse systems with $K_a$ varying from 1 to 100. The solid horizontal line represents the cutoff at 0.01 used previously. The vertical dashed line denotes $s_0^*=3.813$. (b){\it Effective diffusivity in area bidisperse mixtures.} Plot of the effective diffusivity ($D_{eff}$) with respect to increasing area dispersity $\alpha$. Parameter are details provided in Table S2.}
  { \label{fig:highKa}}
\end{figure}

High shape-disparity can amplify coarsening in mixtures, resulting in further enhanced disparity in cell areas. To prevent this coarsening from occuring, we increase the area stiffness $K_a$ to 100. To make sure this does not affect the fluid phase seen in monodisperse mixtures, we study the effective diffusivity as a function of the target shape parameter for several $K_a$ values. We find that $K_a$ barely affects the diffisivities and that the large changes in curvature of $D_{eff}$ versus $s_0$ remain close to $3.81$ such that larger values of $K_a$ do not significantly affect the fluidity of the cells. See Fig.~\ref{fig:highKa}.

\subsection*{Diffusivity of area bidisperse mixtures}
Monodisperse systems with $s_0>3.81$ have a fluid-like diffusivity. Here we check diffusivity for mixtures having the same $s_0=3.85$ for all cell types but bidisperse in size. We see that the average fluid-like diffusivity remains unchanged. See Fig.~\ref{fig:highKa}b.

\subsection*{Component-wise diffusivity and timescales in shape bidisperse mixtures}
\begin{figure}[h]
\centering
\includegraphics[width=\textwidth]{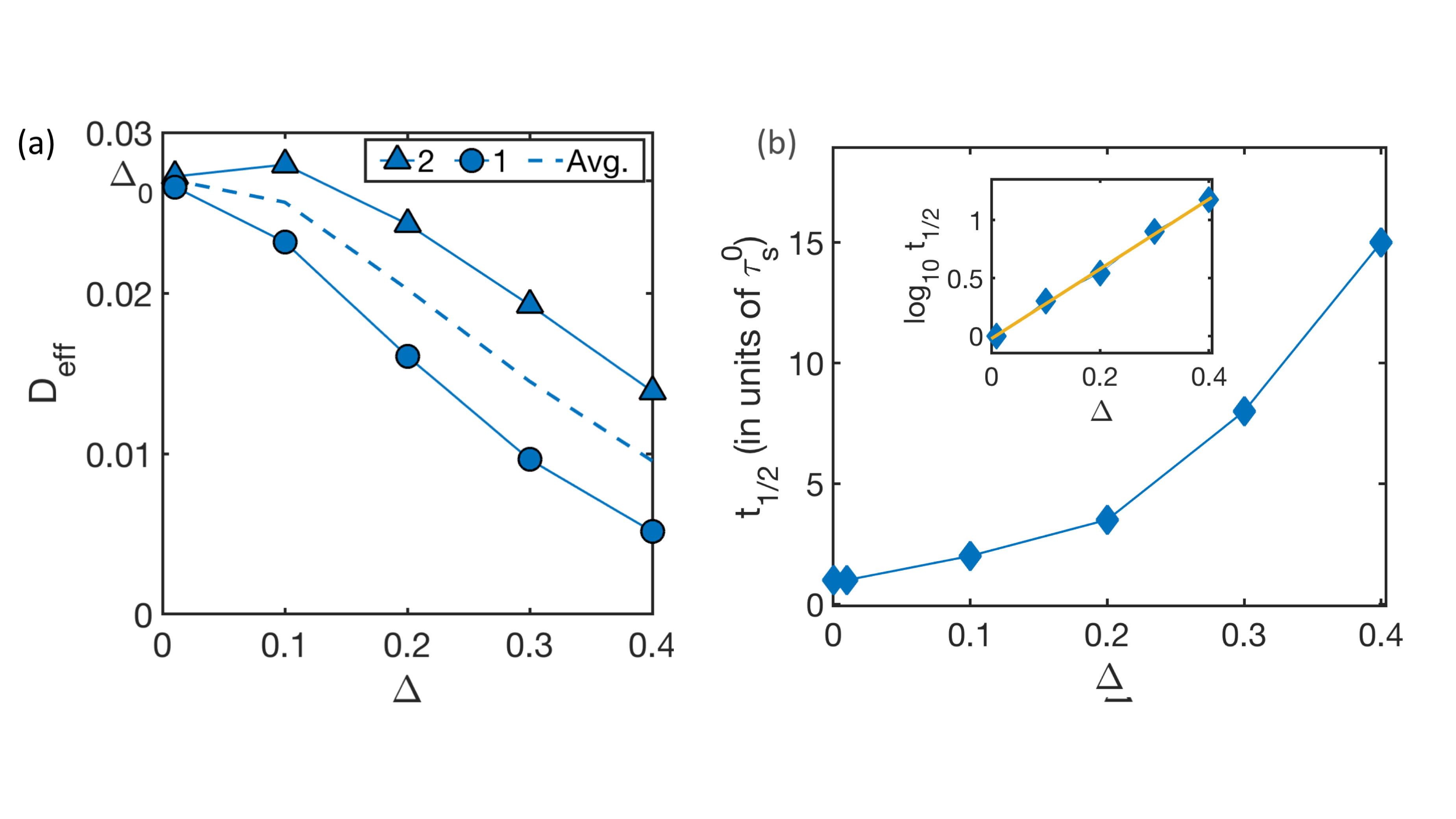}
  \caption{{\it Component-wise diffusivities and timescale to approach steady state.} (a) Plot for effective diffusivity ($D_{eff}$) with respect to increasing shape dispersity $\Delta$. The solid lines are for the two different components with triangles and circles representing higher $s_0$ (type 2) and lower $s_0$ (type 1) respectively. The dashed curve represents the averaged $D_{eff}$. (b) The average time it takes for the system to achieve half of its steady state DP value, or $t_{1/2}$, is plotted against $\Delta$. The solid curves from \ref{fig:shapeG}(b) are used to compute $t_{1/2}$. The inset shows log-log plot of the same, with a linear fit in solid yellow line is $y=3x+3$.}
  { \label{fig:CompDeff}}
\end{figure}

\begin{figure}[h]
\centering
\includegraphics[width=0.58\textwidth]{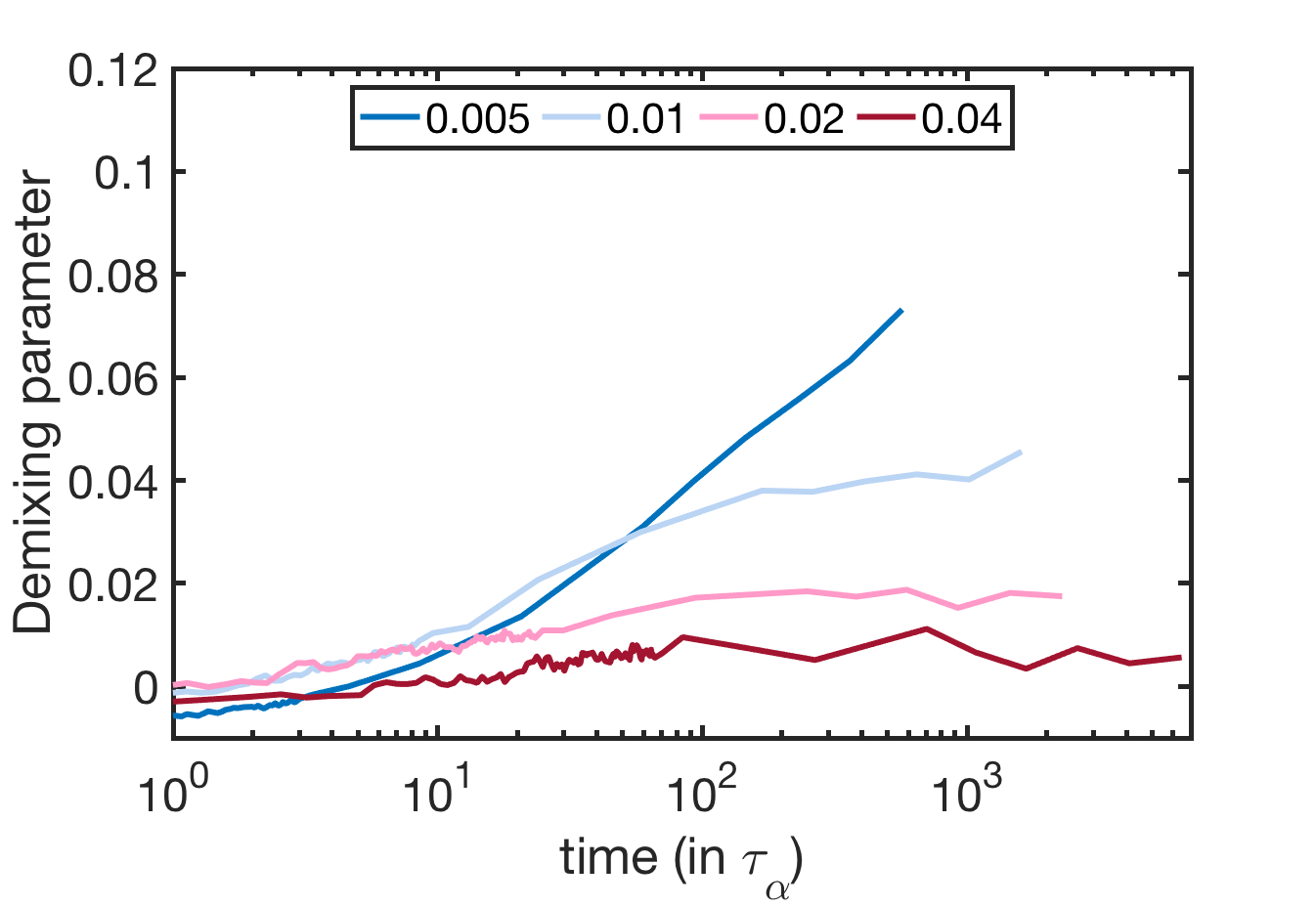}
  \caption{{\it Increasing temperature, diminishes observed demixing. } (a) Plot for demixing parameter ($DP$) with respect to time (in units of $\tau_\alpha$). The solid curves represent an increasing temperature ($T$) from blue ($T=0.005$) to maroon ($T=0.04$). The curves are averaged over 280 ensembles.}
  { \label{fig:DiffTemp}}
\end{figure}

We study the diffusivities of individual components for mixtures with fixed $s_{av}=3.85$. Although increased dispersity signals a solid-fluid mixture, we see that the average behavior remains fluid-like up to high dispersities. Hence, we measure the diffusivity of each component to determine if the solid-like cells diffuse (Fig.~\ref{fig:CompDeff}a). We find that a fluid-like component is indeed able to help the solid-like cells diffuse.

For the demixing observed in shape bidisperse mixtures, as mentioned in the main text, we observe that for most of the $\Delta$, the DP saturates to a final value. We check if the timescale associated with this saturation increases with dispersity since Fig.~\ref{fig:CompDeff}a demonstrates that the solid components (of high dispersity mixtures) do not diffuse as much. We define $t_{1/2}$ as the average time taken by the system to get to half of its final DP. We observe that this half-time increases exponentially with $\Delta$, as shown in the inset of Fig.~\ref{fig:CompDeff}b.

\subsection*{Increasing temperature decreases micro-demixing}
Since we hypothesize that micro-demixing is due to kinetic traps between energy barriers to neighbor exchanges, raising the temperature so that cells should be able to surmount such energy barriers should lead to complete mixing. We, therefore, study the micro-demixing as a function of an increased temperature in a mixture with fixed dispersity $\Delta=0.2$. We observe that increasing temperature indeed leads to complete mixing, i.e. the demixing parameter goes to zero. For an increased temperature we use the relation $\tau_\alpha\propto T^{-3/2}$, reported in ~\cite{Sussman2018c}, to re-scale the x-axis. The lower temperatures systems have yet to reach a steady state demixing value. One can use a strip geometry to probe the exact steady state value for lower temperatures, which we leave for future work.

\subsection*{Cortical tension for sorted vs. mixed configurations}

An emergent line tension between two different kinds of cells must show a high line tension along heterotypic edges and lower line tension along the homotypic edges. Hence, for both the sorted and mixed scenarios (Fig.~\ref{fig:cost}), we study a \textit{line tension map} where the thickness of the edge is linearly proportional to its line tension. A positive value is colored in red and a negative value is colored in blue. The cortical tension for each edge can be computed using the method suggested in Ref.~\cite{Yang12663}.

 \begin{figure}[!htbp] 
\centering
\includegraphics[width=0.99\textwidth]{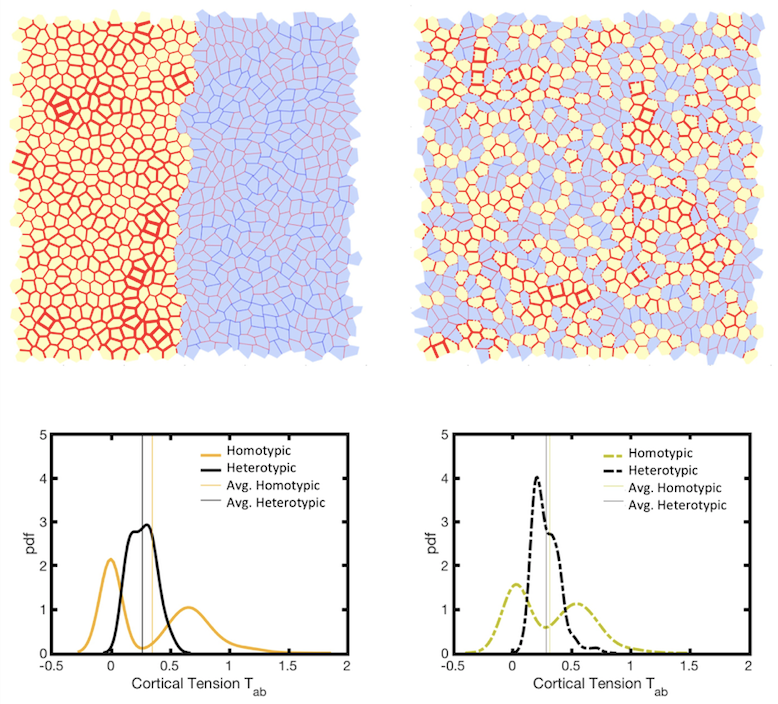}
  \caption{{\it Cortical tension in shape bidisperse mixtures of $\Delta=0.4$.} Left and right panels shows line tension maps for sorted and mixed scenarios for a $N=900$ system respectively. Heterotypic edges are shown in dash-dot lines. Yellow and blue cells have $p_0=3.65$ and $4.05$ respectively. They are followed by histograms for heterotypic(in black) and homotypic(colored) edges. Vertical lines show the mean values for each curve in their respective colors.}
  { \label{fig:shapeTension}}
\end{figure}

 \begin{figure}[!htbp] 
\centering
\includegraphics[width=0.99\textwidth]{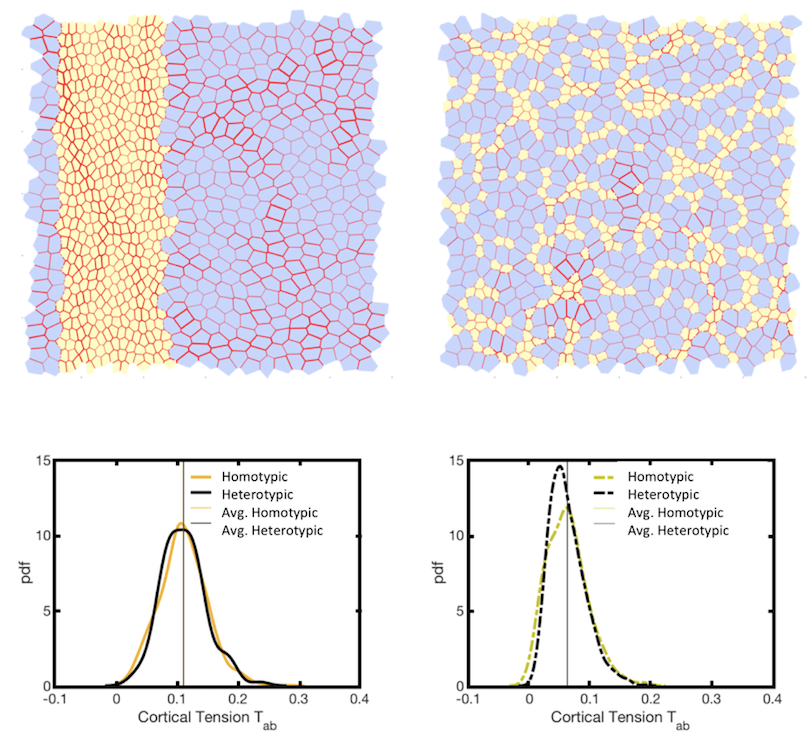}
  \caption{{\it Cortical tension in area bidisperse mixtures of $\alpha=2.5$.} Left and right panels shows line tension maps for sorted and mixed scenarios for a $N=900$ system respectively. Heterotypic edges are shown in dash-dot lines. Yellow and blue cells have $A_0=0.57$ and $1.43$ respectively. They are followed by histograms for heterotypic(in black) and homotypic(colored) edges. Vertical lines show the mean values for each curve in their respective colors.}
  { \label{fig:areaTension}}
\end{figure}

 The cortical tension analysis conveys the fact that there is no emergent line tension due to bidispersity in the mixtures we study. The mean heterotypic line tension (black vertical line) is less than or equal to the mean homotypic line tension (colored vertical line) for all the scenarios. See Figs.~\ref{fig:shapeTension} and ~\ref{fig:areaTension}.

\subsection*{Differential T1 energy barriers in area bidisperse mixtures}

\begin{figure}[hbt] 
\centering
\includegraphics[width=0.99\textwidth]{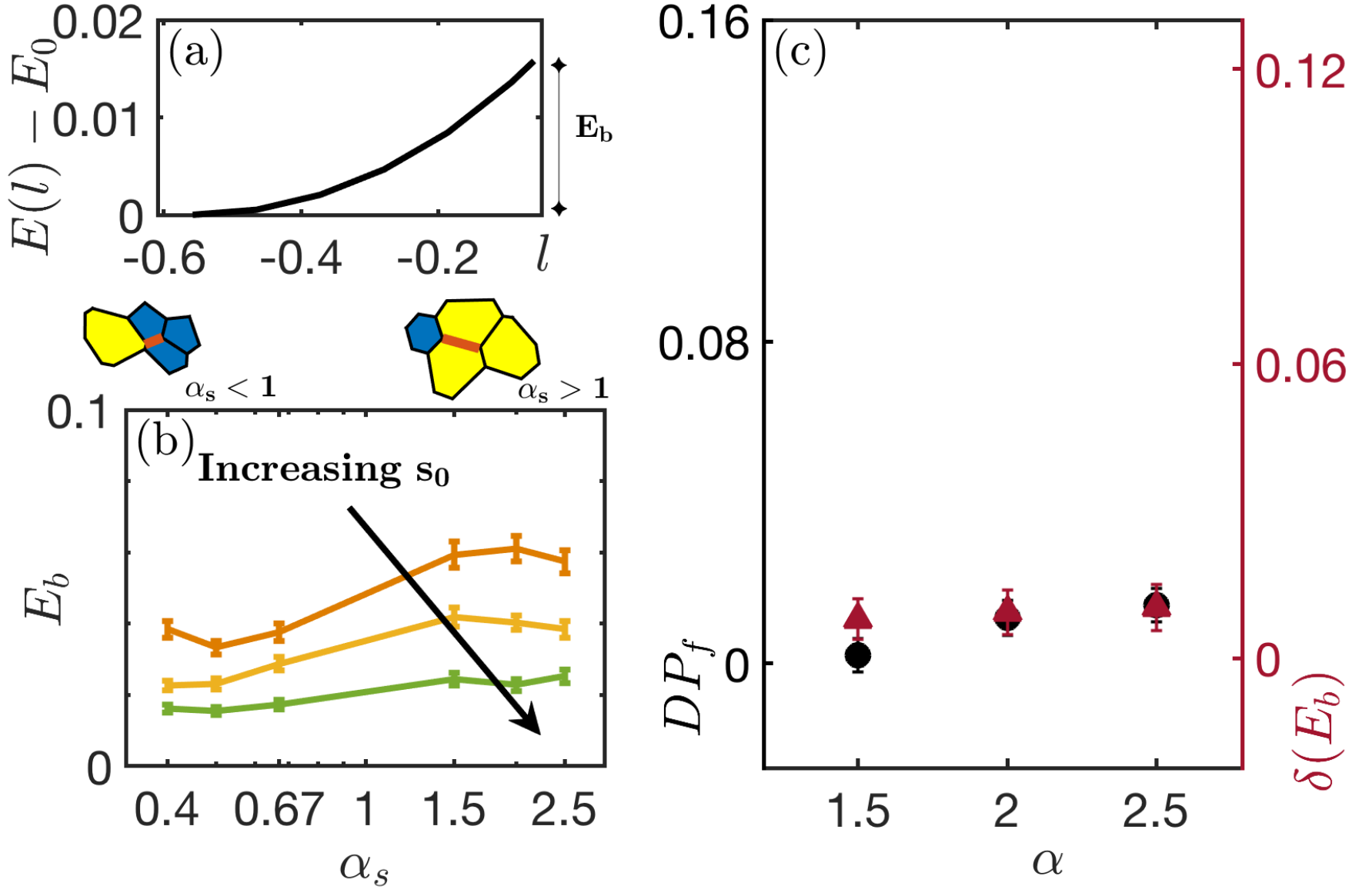}
  \caption{{\it Differential energy barriers in area bidisperse mixtures.} (a) Energy $E(l)$ relative to $E_0$ versus T1 edgelength $l$ for a typical size bi-disperse T1 pair ($\alpha=2.5, s_{av}=3.85$). (b) Energy Barrier $E_b$ is plotted against area disparity $\alpha_s$ where large values on right and small values on left imply large-cell cluster in yellow and small-cell cluster in blue respectively. Each solid curve represents the barrier for a heterotypic cell to get out of the cluster for a fixed $s_{0}$ (varied from solid-like (orange) to liquid-like (green) - 3.82,3.85,3.88) (c) Correlation plot for $s_{0}=3.85$ between Differential Energy Barriers on the right y-axis $\delta(E_b)$ (in maroon triangles) and demixing relative to mixed scenario $DP_f$ on the left y-axis (in black discs). Size ratio $\alpha$ is plotted on x-axis. Simulation details provided in Table S4.
  }
\label{fig:corr}
\end{figure}

We present data supporting the notion that the differential energy barriers are smaller for the area bidisperse mixtures as compared to the shape bidisperse ones. We focus on larger cells trying to invade a cluster of smaller cells and vice versa to determine the stability (or lack thereof) an interface. See Fig.~\ref{fig:corr}. We also present data for other types of topologies for both shape and area bidisperse mixtures for completeness (see Fig.~\ref{fig:otherT1s}). 

Finally, to study differential T1 energy barriers in a simplified setting, we consider four cells connected to each other symmetrically. The energy is minimized with respect to a diminishing T1 edge length $l$ using MATLAB. The area stiffness is kept very high and the initial condition is recursively fed from a longer $l$ energy minimized configuration to the subsequent shorter $l$. We can accommodate different sizes and shapes as long as cells of different types are positioned symmetrically about both $x$ and $y$ axis and make the cells sharing the T1 edge (T1 pair) have different properties from the non-T1 pair. The formula used to compute energy barrier is  $E(l=l_{H})-E(l=0)$, where $l_{H}$ is the edge length of a uniform hexagon with unit area.

 \begin{figure}[!htbp] 
\centering
\includegraphics[width=0.8\textwidth]{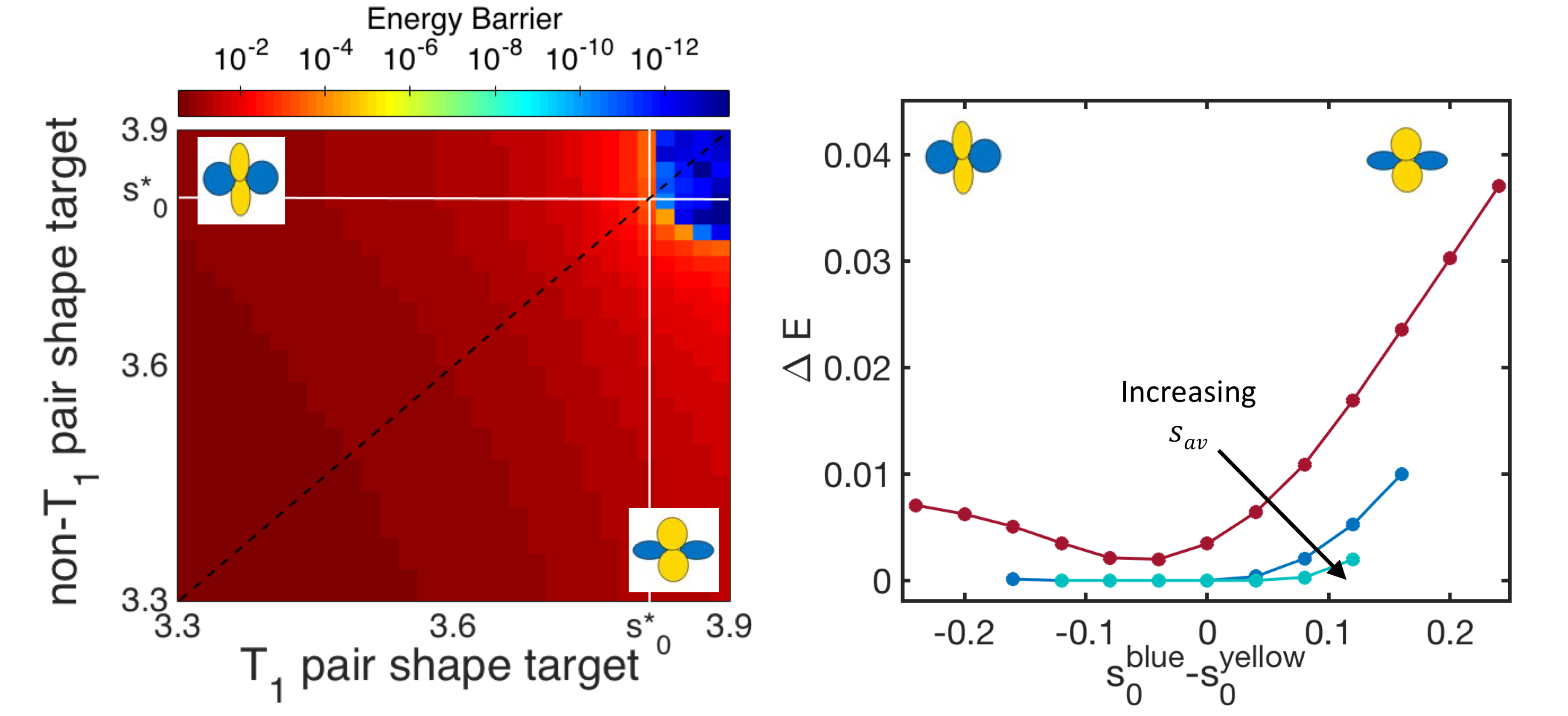}
  \caption{{\it Symmetric 4-cell T1 energy barriers for shape bidispersity.} On the left is the color plot of energy barrier as a function of independently tunable shapes of T1 pair and non-T1 pair is plotted along x-axis and y-axis respectively. The dashed line represents monodisperse calculation ie for $\Delta=0$. As expected it is red till it reaches the monodisperse transition point $s*_0=3.813$, after which it becomes blue. Off-diagonal phase points depict bidisperse mixtures ie $\Delta\neq0$. We see that it is necessary for the T1-pair to be fluid like, for vanishing barrier. On the right is a cross-section of the phase diagram on left. Energy barrier is plotted against area disparity for increasing values of $s_0$ 3.79 to 3.85.}
  { \label{fig:4cellShape}}
\end{figure}

 \begin{figure}[!htbp] 
\centering
\includegraphics[width=0.99\textwidth]{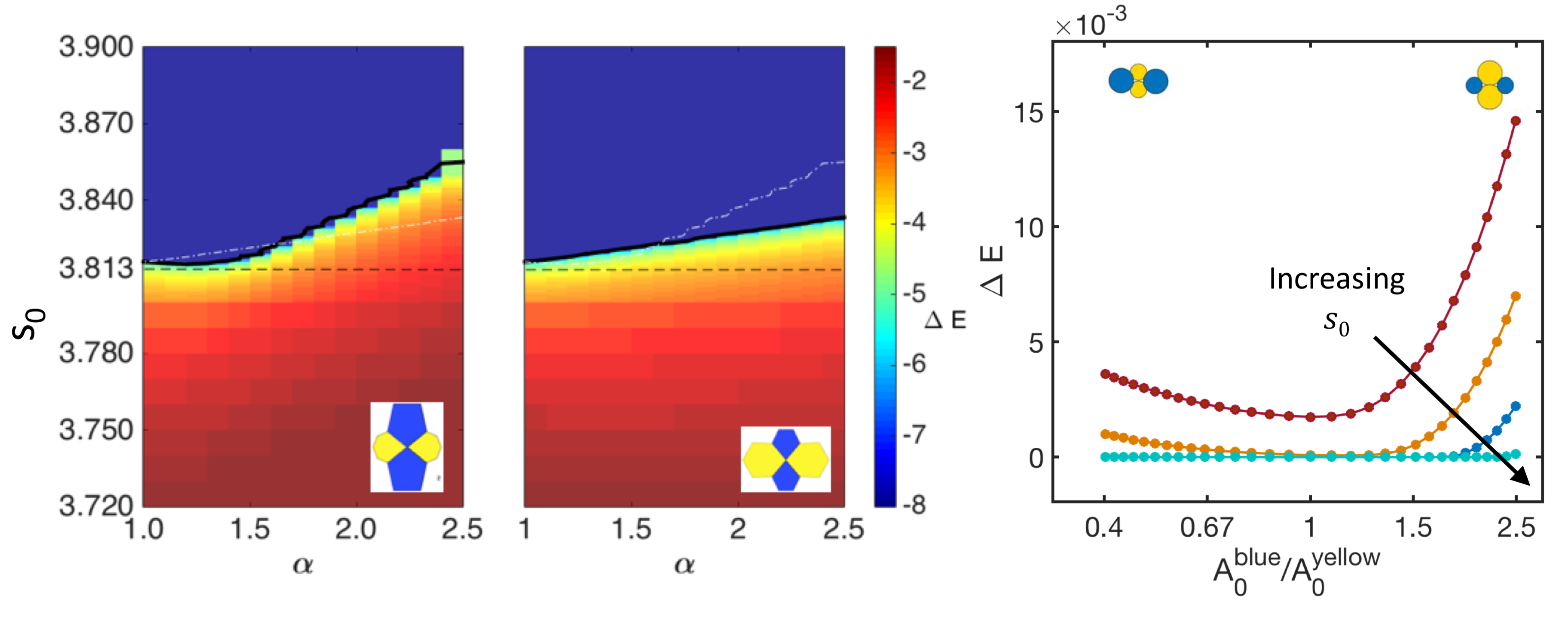}
  \caption{{\it Symmetric 4-cell T1 energy barriers for area bidispersity.} On the left is the color plot of energy barrier as a function of independently tunable sizes of T1 pair (blue polygons) and non-T1 pair(yellow polygons). On the left is when blue polygons are bigger than yellow. On the right is smaller blue cells sandwiched between yellow (BssB). $\alpha$ and $P_0$ are the area ratios and preferred shape index respectively. Dashed black line represents the monodisperse transition point $s_0^*=3.813$. This graph predicts the energy barriers to vanish at a shape index higher than $s_0^*$ in highly bidisperse systems. On the right is a cross-section of the cumulative phase diagram on left. Energy barrier is plotted against area disparity for increasing values of $s_0$ 3.79 to 3.85.}
  { \label{fig:4cellarea}}
\end{figure}

To study the effect of shape bidispersity (Fig.~\ref{fig:4cellShape}), the energy barrier (red when non-zero and blue when vanishingly small) is plotted with respect to the shape of T1 pair (x-axis) and the shape of the non-T1 pair (y-axis), which can be independently varied. A similar analysis is done for mixtures with bidisperse areas (Fig.~\ref{fig:4cellarea}). We observe differential energy barriers in both cases with, again, the size of the barrier generically larger in the shape bidisperse case as compared to the area bidisperse case, even in this simplified calculation. Re-phrasing this in terms of invading a cluster of the opposite type, one can think of these as invading doublets of opposite kind.

 \begin{figure}[!htbp] 
\centering
\includegraphics[width=\textwidth]{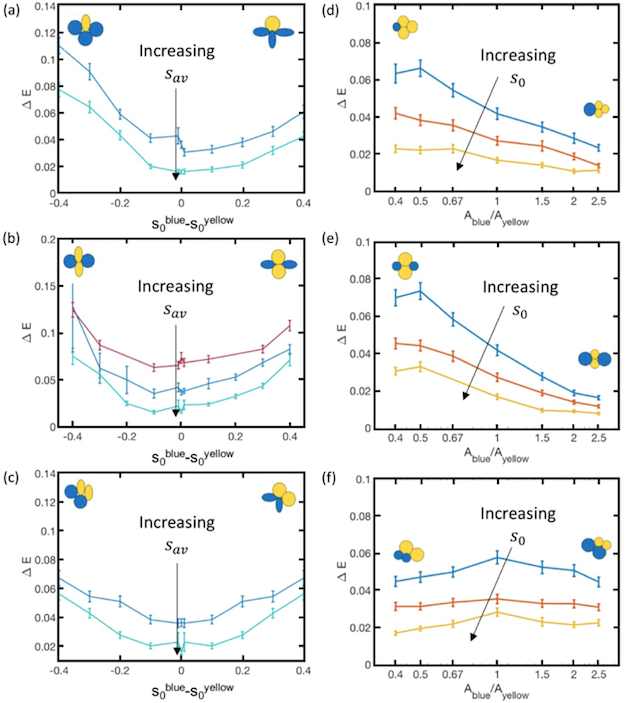}
  \caption{{\it T1 transitions in shape and area bidisperse mixtures.} (a)-(c) T1 topologies (shown as cartoons on axis extremities) and their barrier statistics for shape bidisperse mixtures. (d)-(f) T1 topologies (shown as cartoons on axis extremities) and their barrier statistics for size bidisperse mixtures. Parameters used are in Table~\ref{tab:T1FIRE}. }
  { \label{fig:otherT1s}}
\end{figure}

\subsection*{Additional experimental features}

Figure S10 provides additional information regarding the motility of the cells (Fig. S10a), the distribution of cell areas (Fig. S10b), and the ratio of the numbers of the two cell types during the course of the experiment (Fig. S10c).  There is little difference in the amount of the displacement the two different cell types undergo within 24 hours either in the monotypic monolayers or in the combined monolayers. There is also minimal difference in the distribution of cell areas for the two different cell types as measured in the respective monotypic monolayer (Fig. S10b). FInally, Fig. S10c demonstrates that the Ctr-E-cad-$^{-/-}$ mixtures remain approximately 50:50 mixtures over the duration of the experiments.

We verify that natural variability in adhesion from cell-to-cell does not drive micro-demixing by calculating the demixing parameter (DP) for monotypic monolayers with half the cells tagged with one type of stain and the other half tagged with a second stain (Fig. S11a).  We find that the demixing parameter does not increase (or decrease) on average with time, strongly suggesting that differential adhesion is indeed what is driving the micro-demixing. 

In addition to computing the demixing parameter for the co-culture, we also study the pair correlation functions between all the three possible cell-type pairs in the co-culture in the high calcium condition (Figs. S11b and c). As the demixing parameter value is rather consistent with the prediction, the pair correlation function also indicates a small-scale correlation across a couple of cell diameters. The experimental pair-correlation curve is more structureless than the predicted curve, which potentially can be understood given the variability in cell areas found for both the cell-types shown in Fig. S11c.

 \begin{figure}[!htbp] 
\centering
\includegraphics[width=\textwidth]{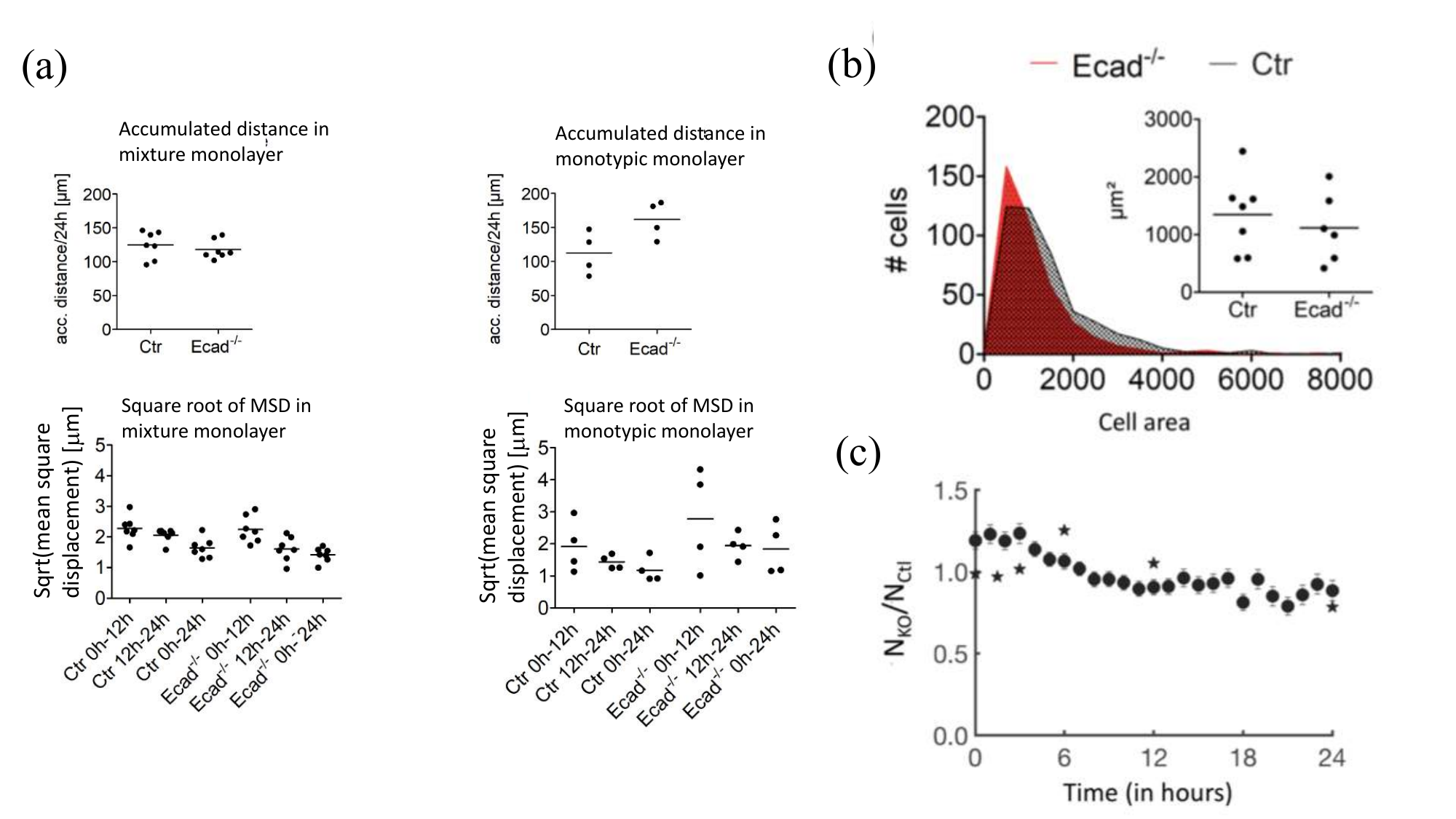}
  \caption{{\it Additional quantification of cell properties.} (a) Cell displacements integrated over 24 hours in the Ctr-E-cad$^{-/-}$ mixtures and in the control Ctr-Ctr and E-cad$^{-/-}$-E-cad$^{-/-}$ monotypic monolayers (cells of the same type but different tag). Ten cells from each type of the 7 Ctr-E-cad$^{-/-}$ demixing videos and the 4 Ctr-Ctr and Ecad$^{-/-}$-Ecad$^{-/-}$ demixing videos are measured. We additionally show the root mean-square displacements for various time intervals for the same data sets. (b) The distribution of cell areas is shown for E-cad$^{-/-}$ (in red) and Ctr (in black). The inset shows the average across 6 and 7 isolates respectively, to show that area distributions are similar for both cell-types. (c) The ratio of- number of E-cad$^{-/-}$ (KO) to Ctr cells  is plotted against time for a typical experimental co-culture. The ratio approaches unity i.e. it is almost a 50:50 mixture, over the course of the experiment. }
  { \label{fig:AaronSI}}
\end{figure}

 \begin{figure}[!htbp] 
\centering
\includegraphics[width=0.9\textwidth]{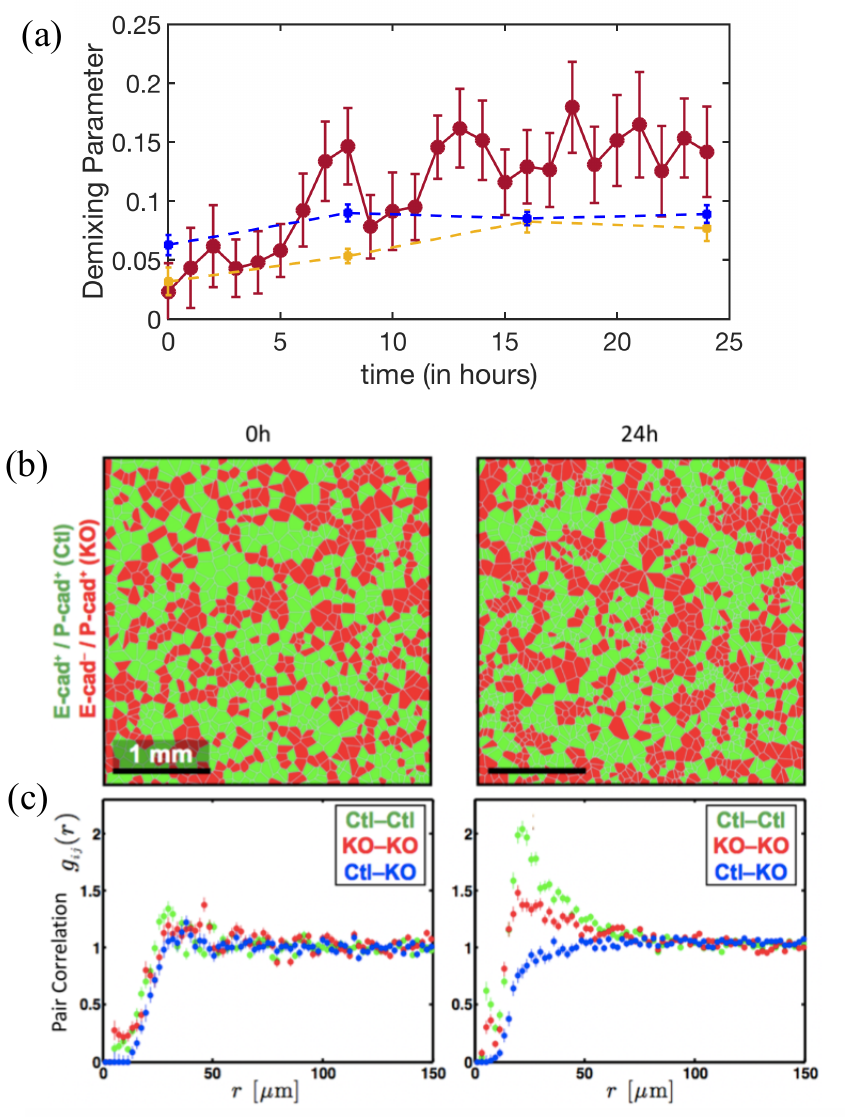}
  \caption{{\it Additional quantification of experimental co-cultured monolayers.} (a) The solid maroon curve represents the time evolution of the demixing parameter for the E-cad$^{-/-}$ cell-type in the Ctr-E-cad$^{-/-}$mixture as a function of time, averaged over 5 different monolayers using five different isolates. The maroon curve should be compared against the almost flat demixing curves for monotypic mixtures composed of 50:50 differently tagged Ctr and E-cad$^{-/-}$  cells shown in yellow and blue dashed curves respectively, averaged over two monolayers using two different isolates each. (b) The initial (0h) Voronoi tessellation of the co-culture nuclei, is compared side by side to the final (24h) snapshot in the high calcuim condition. Green cells and red cells depict Ctr (Ctl) and E-cad$^{-/-}$ (KO) cells respectively. (c) The pair correlation function is plotted for the initial and final snapshots in (b). Green, red and blue markers depict correlations between- homotypic Ctl, homotypic KO, and heterotypic Ctl-KO nuclei respectively. }
  { \label{fig:AaronSI}}
\end{figure}

\end{document}